\documentclass[11pt,a4paper]{article}
\pdfoutput=1
\usepackage{colortbl}
\usepackage{jheppub}

\usepackage{epsfig,multicol,bbm}
\usepackage{amsmath,amsfonts,amssymb,longtable,mathtools}
\usepackage{array}
\usepackage{graphicx}
\usepackage{subfig}
\usepackage{enumerate}
\usepackage[dvipsnames]{xcolor}
\usepackage{placeins}
\usepackage{longtable,booktabs,caption}
\usepackage{changepage}
\usepackage{float}
\newcommand{\df}{\dot{\phi}}
\newcommand{\ddf}{\ddot{\phi}}

\def\bea{\begin{eqnarray}}
\def\eea{\end{eqnarray}}

\def\Mp{M_{\rm Pl}}

\input epsf.sty

\usepackage[makeroom]{cancel}
\usepackage{latexsym}
\usepackage{epsf}
\usepackage{amssymb}
\usepackage{graphicx}
\usepackage{amsmath}
\usepackage{amsmath,amssymb,amsthm}
\usepackage{verbatim}
\usepackage{hyperref}
\renewcommand{\d}{\textrm{d}}

\def\ni{\noindent}
\def\bi{\begin{itemize}}
\def\ei{\end{itemize}}
\def\be{\begin{equation}}   
\def\ee{\end{equation}}
\def\ben{\begin{equation*}}
\def\een{\end{equation*}}
\def\c1{C_1}
\def\d1{D_1}

\DeclareMathOperator{\Tr}{Tr}

\preprint{UTTG-20-2021}

\title{Rapid-turn inflation in supergravity is rare and tachyonic}

\author[a]{Vikas Aragam,}
\author[b]{Roberta Chiovoloni,}
\author[a]{Sonia Paban,}
\author[a]{Robert Rosati,}
\author[b]{Ivonne Zavala}

\affiliation[a]{Department of Physics, University of Texas at Austin, Austin, TX 78712, USA}
\affiliation[b]{Department of Physics, Swansea University, SA2 8PP, UK}

\emailAdd{aragam@utexas.edu}
\emailAdd{r.chiovoloni.967740@swansea.ac.uk}
\emailAdd{paban@physics.utexas.edu}
\emailAdd{rjrosati@utexas.edu}
\emailAdd{e.i.zavalacarrasco@swansea.ac.uk}

\abstract{Strongly non-geodesic, or rapidly turning trajectories in multifield inflation have attracted much interest recently from both theoretical and phenomenological perspectives. Most models with large turning rates in the literature are formulated as effective field theories. In this paper we investigate rapid-turn inflation in supergravity as a first step towards understanding them in string theory.  We find that large turning rates can be generated in a wide class of models, at the cost of high field space curvature. In these models, while the inflationary trajectories  are stable,  one Hessian eigenvalue is always tachyonic and large, in Hubble units. Thus, these models satisfy the  de Sitter swampland conjecture along the inflationary trajectory. However, the high curvatures underscore the difficulty of obtaining rapid-turn inflation in realistic string-theoretical models.
In passing, we revisit the $\eta$-problem in multifield slow-roll inflation and show that it does not arise, inasmuch as the inflatons, $\phi^i$, can {\em all} be heavier (in absolute value) that the Hubble scale: $|m_i|/H>1$, $\forall\, i$.  }


\begin{document}
\maketitle

\section{Introduction}
Cosmological inflation is the leading mechanism for explaining the origin of primordial perturbations, which seeded the large-scale structures that we observe today.  The most recent cosmological observations are consistent with the simplest inflationary scenario, in which the potential energy of a single scalar field drives a period of early cosmological acceleration \cite{2020}. However, multifield models of inflation with strongly non-geodesic trajectories have received recent interest, motivated by theoretical and phenomenological constraints.
Strongly non-geodesic, or rapidly turning inflationary trajectories can satisfy the recently proposed consistency conjectures on the inflationary scalar potential \cite{Hetz:2016ics,Achucarro:2018vey,Obied:2018sgi,Garg:2018reu,Ooguri:2018wrx}. 
Moreover, rapid-turn models can admit fat fields, which are heavier than the Hubble scale and thus avoid the $\eta$-problem \cite{Chakraborty:2019dfh}. 
Phenomenologically, strongly non-geodesic motion can have interesting observational implications, such as breaking the single-field consistency relations between observables \cite{Kaiser:2013sna,Christodoulidis:2018qdw,Christodoulidis:2019hhq,Christodoulidis:2019jsx}, leaving signature features in the primordial powerspectrum \cite{Chen:2018brw,Chen:2018uul,Slosar:2019gvt,Braglia:2021ckn,Braglia:2021rej}, producing primordial black holes \cite{Fumagalli:2020adf,Braglia:2020eai,Palma:2020ejf,Anguelova:2020nzl}, and sourcing a stochastic background of gravitational waves \cite{Barausse:2020rsu,Fumagalli:2020nvq,Fumagalli:2021cel,Domenech:2021ztg}.

In an effective derivative expansion, multifield physics is characterised by a field space metric, a potential, and couplings to the spacetime metric via the Ricci scalar. The true multifield nature of the trajectory manifests when the trajectory strongly deviates from geodesic motion in field space. This deviation is measured by the dimensionless turning rate $\Omega/H$, where $H$ is the Hubble parameter. 
For minimally coupled scalars, one finds \cite{Hetz:2016ics,Achucarro:2018vey} that along solutions to the equations of motion,
\begin{equation}
 \epsilon_V= \epsilon \left\{ \left(1 + \frac{\eta}{2 (3-\epsilon)}\right)^2 + \frac{\Omega^2}{9 H^2} \frac{1}{(1-\epsilon/3)^2} \right\},
 \label{eq:Ev_Eh_relationship}
\end{equation}
where the slow-roll and gradient parameters are defined as:
\begin{align}
	\epsilon \equiv -\frac{\dot{H}}{H^2}, \quad \eta \equiv \frac{\dot{\epsilon}}{H\epsilon}, \quad \epsilon_V \equiv \frac{\Mp^2}{2} \frac{\left| \nabla V \right|^2}{V^2}.
\end{align}

Slow-roll inflation, with $\epsilon \ll 1$ and $\eta \ll1$, can happen in different regions of the $\Omega/H $ and $\epsilon_V $ plane. Potentials and metrics for which $\epsilon_V \ll 1$ allow for an effective single-field, slow-roll trajectory, whereas steep potentials ($\epsilon_V \gtrsim 1$) can  feature rapid-turn, slow-roll trajectories ($\Omega/H \gg1$, $\epsilon \ll 1$)  \cite{Hetz:2016ics,Achucarro:2018vey}. The conditions for rapid-turn, slow-roll trajectories are known in the two-field case \cite{Bjorkmo:2019fls, Bjorkmo:2019aev,Chakraborty:2019dfh} and will be reviewed in Section \ref{sec1}. The conditions for three or more fields in the low-torsion limit\footnote{See also \cite{Pinol:2020kvw} for a discussion on multifield inflation with any number of fields.} are discussed in \cite{Aragam:2020uqi}. 

Moreover, it was shown in  \cite{Chakraborty:2019dfh} that a sufficient condition for strongly non-geodesic trajectories is to have all fields heavier than the Hubble scale, i.e. {\em fat inflation}. 
This is a remarkable result, as it effectively evades the $\eta$-problem in multifield inflation.
 As we show in section \ref{sec1}, more generally, multifield inflation does not have an $\eta$-problem, inasmuch as all the inflations can be heavier than the Hubble parameter ($|m_i|/H\gg 1$).  
In \cite{Chakraborty:2019dfh} it was also shown that  fat inflatons are sufficient, but not necessary for rapid-turn inflation. 
An example of this is orbital inflation \cite{Achucarro:2019mea}, which produces rapid-turn inflation with a large negative mass-squared along the inflationary trajectory \cite{Aragam:2019omo} (se also \cite{Achucarro:2019pux})
Note that the  tachyons along the inflationary trajectory do not imply an instability\footnote{This is standard for concave potentials, such as in single field Starobinsky inflation \cite{Starobinsky:1980te}.}. Indeed,  the trajectory can be stable in that the  Lyupanov exponents contain a compensating term from the turn rate \cite{Christodoulidis:2019hhq,Christodoulidis:2019mkj}. 

Multifield models of inflation are not inevitable, but they arise naturally in supergravity and string theory compactifications. Thus, it is sensible to ask how common is strongly non-geodesic, slow-roll behaviour in supergravity, being a low energy effective theory description of string theory. 
In this paper we aim at addressing precisely these two questions. 

The rest of the paper is organised as follows. In Section \ref{sec1}, we review the conditions for  slow-roll inflation in the multifield case and show that the $\eta$-problem does not arise. That is, the masses of all the inflaton fields do not need to be smaller than the Hubble parameter for a sustained period of inflation when more than one field is present. In Section \ref{subsec1}, we discuss  possible two-field trajectories in detail, and apply these results to the supergravity case in Section \ref{sec2}. In this section, we show how to construct rapid-turn trajectories in a large class of supergravity models. In these models, the inflaton superfield does not mix with the supersymmetry breaking direction, i.e. the inflaton and Goldstino superfields are orthogonal. 
In Appendix \ref{sec:survey}, we survey a broad set of models in the literature, including several that do not belong to this class. We find that the non-orthogonal models have neither large turning rates nor fat fields.
In Appendix \ref{sec:sGI}, we analytically explore the possibility of large turning rates in single superfield inflation. A survey of this class of models is also included in Appendix \ref{sec:survey}.
 
%%%%%%
We find that the surveyed models with large turning rates, as well as all rapid-turn models we construct, have a large (w.r.t.~the Hubble scale) tachyonic direction along the inflationary trajectory. Thus, these models satisfy the swampland de Sitter conjecture (dSSC) \cite{Garg:2018reu,Ooguri:2018wrx} along the inflationary trajectory.
This inflationary regime only occurs when one of the coefficients in the K\"ahler potential is unnaturally small and the field space curvature is high. We therefore argue that rapid-turn inflation is rare in supergravity.  We finish with a summary and discussion of our results in Section \ref{sec3}.

\section{Slow-roll Multifield Inflation}\label{sec1}

In this section we review slow-roll, multifield inflation following \cite{Chakraborty:2019dfh,Aragam:2020uqi}. 
The  Lagrangian for several scalar fields minimally coupled to gravity takes the form:
\be\label{4Daction}
S= \int{d^4x \sqrt{-\tt g} \left[\Mp^2 \frac{R}{2}  - \frac{g_{ab}(\phi^c)}{2} \partial_\mu\phi^a \partial^\mu\phi^b - V(\phi^a)\right]} \,,
\ee
where $g_{ab}$ is the field space metric, $\Mp^2=(8\pi G_N)^{-1}$ is the Planck mass, and $G_N$ is Newton's constant. In an FLRW spacetime, the equations of motion are
\bea
&&H^2 = \frac{1}{3\Mp^2} \left(\frac{\dot \varphi^2}{2}  + V(\phi^a)\right) \,, \label{H} \\
&& \ddot \phi^a + 3H\dot\phi^a + \Gamma^a_{bc} \dot\phi^b\dot\phi^c + g^{ab} V_{b} =0  \,,
\label{phis}
\eea
where 
\be\label{varphi}
\dot\varphi^2 \equiv g_{ab} \dot \phi^a\dot\phi^b\,.
\ee
The Christoffel symbols in \eqref{phis} are computed using the scalar manifold metric $g_{ab}$ and $V_a$ denotes derivatives with respect to the scalar field $\phi^a$.

\subsection{Kinematic basis decomposition}

The slow-roll conditions and rapid turning in multifield inflation  can be understood neatly by using a  kinematic basis to decompose the inflationary trajectory into tangent and normal directions.
Let us introduce unit tangent and normal vectors to the inflationary trajectory, $T^a$ and $N^a$, as follows:
%%%
\be
T^a = \frac{\dot\phi^a}{\dot\varphi}\,, \qquad T^aT_a =1\,,\qquad
N^aT_a=0, \qquad N^aN_a=1\,.
\ee
This basis is useful in studying the slow-roll conditions, which mimic the single field case. 
Projecting the equation of motion \eqref{phis} for the scalars $\phi^a$ along these two directions gives:
\bea
 \ddot\varphi + 3H\dot\varphi + V_T & = &0 \,, \label{varphiT}\\
D_t T^a  + \Omega N^a \label{Omega1} &=&0\,,\label{varphiN}
\eea
 where $V_T = V_aT^a$, $V_N = V_aN^a$ and the {\em turning rate} parameter $\Omega$ is defined as
\be\label{Omega}
\Omega \equiv \frac{V_N}{\dot\varphi}\,.
\ee 
The field-space covariant time derivative is  defined as:
 \be\label{Dt}
 D_tT^a \equiv \dot T^a + \Gamma^{a}_{bc} T^b \dot \phi^c \,.
 \ee
To study the masses of the scalar field, we introduce the following matrix:
\be
{\mathbb M}^a_{\,\,b}\equiv \Mp^2\frac{ \nabla^a \nabla_b V}{V}, 
\ee
where $\nabla_a A_b \equiv \partial_a A_b - \Gamma_{ab}^{c}A_c$, for some vector $A_b$. For two fields, this can be written as
\begin{equation*}\label{Hessian}
   {\mathbb M} = \frac{\Mp^2}{V}\left(
      \begin{array}{cc}
    V_{TT} & V_{TN}  \\
        V_{NT} & V_{NN} 
           \end{array} \right),
  \end{equation*}
where  $V_{TT} = T^a T^b \nabla_a \nabla_b V$, etc., and we can now define the parameter $\eta_V$ as:
\be\label{etaV}
\eta_V \equiv \left|{\rm min\,\,\, eigenvalue} \{\mathbb M\}\right|\,.
\ee
Note also that the eigenvalues of ${\mathbb M}$ in the two field case can be written neatly as
\be\label{eigenvalues}
\lambda_\pm = \frac{1}{2} \left({\rm Tr\,} {\mathbb M} \pm \sqrt{ {\rm Tr\,}  {\mathbb M}^2 - 4\, {\rm det}  {\mathbb M}}     \right).
\ee
For example, if all eigenvalues are positive as in the case of fat inflation, one has  that $0<{\rm det}\,{\mathbb M}\leq {\rm Tr}\,{\mathbb M}^2/4$.

\smallskip

We now summarize useful expressions for the tangent and normal projections of the mass matrix elements.
Taking the time derivative of eq.~\eqref{varphiT}, we obtain an expression for the tangent projection, that is \cite{Achucarro:2010da,Hetz:2016ics,Christodoulidis:2018qdw,Chakraborty:2019dfh}: 
\be\label{VTT1}
\frac{V_{TT}}{3H^2} =  \frac{\Omega^2}{3H^2} + \epsilon -\delta_\varphi -
\frac{\xi_\varphi}{3}   \,,
\ee
where we have introduced the {\em slow-roll} parameters:
\bea
\epsilon\equiv -\frac{\dot H}{H^2} &=& \frac{\dot\varphi^2}{2M^2_{Pl}H^2} \,,\label{epsilonH}\\
\delta_\varphi\equiv \frac{\ddot{\varphi}}{H \dot{\varphi}} \,,\label{etaphi} \\
\xi_\varphi \equiv \frac{\dddot\varphi}{H^2 \dot\varphi}\,.
\eea

\ni Next, taking the time derivative of eq.~\eqref{Omega}, we obtain an expression for $V_{TN}$ as \cite{Achucarro:2010da,Hetz:2016ics}: 
\be\label{VTN1}
\frac{V_{TN}}{H^2} = \omega\left( 3-\epsilon+2 \,\delta_\varphi +\nu \right),
\ee
where we introduced the dimensionless turning rate $\omega\equiv \Omega/H$ and its fractional derivative
\be\label{nu}
 \nu\equiv \frac{\dot\omega}{H\omega}\,.
 \ee
Note that these relations are exact, as we have not made use of any slow-roll approximations. 
We observe that $V_{TT}$ and $V_{TN}$ can be written in terms of the turning rate and the slow-roll parameters. On the other hand, $V_{NN}$ depends on the inflationary trajectory in a model-dependent manner. 

\subsection{Slow-roll in multifield inflation}

A nearly exponential expansion is ensured by requiring the fractional change of the Hubble parameter per e-fold, $d(\ln H)/d{ N}$ to be small. This corresponds to $\epsilon \ll1$. In order for inflation to last for a sufficiently long time and solve the horizon problem, one also requires that $\epsilon$ remains small for a sufficient number of Hubble times. This is measured by the second slow-roll parameter $\eta$, defined as:
\be\label{etaH}
\eta \equiv \frac{\dot \epsilon}{\epsilon H}  = \frac{\ddot{H}}{H \dot{H}}+2\epsilon 
= 2 \,\delta_\varphi + 2\,\epsilon \ll 1 \,,
\ee    
Since $\epsilon \ll 1$ in slow-roll, a small $\eta$ requires $\delta_\varphi \ll 1$.
 
Using the Friedmann equation, one can see that the first slow-roll condition, $\epsilon\ll1$, implies that $\dot\varphi^2\ll V$. Hence we can write
\be\label{Hslow}
H^2 \simeq \frac{V}{3M_{Pl}^2} \,.
\ee
Moreover, $\delta_\varphi \ll 1$ implies that we can write \eqref{varphiT} as
\be\label{phislow}
3H \dot\varphi + V_T \simeq 0 \,.
\ee
Using  \eqref{phislow} and \eqref{Hslow}, these conditions further imply
\be\label{epT}
\epsilon_T \equiv \frac{M_{Pl}^2}{2} \left(\frac{V_T}{V}\right)^2 \ll1  \,,
\ee
such that only the tangent projection of the derivative of the potential need be small.
Note  that during slow-roll, we can write \eqref{Omega} as
\be\label{Omega2}
\frac{\Omega}{H} \simeq -3 \frac{V_N}{V_T} \,.
\ee
Additionally, from \eqref{VTT1} we see that during slow-roll,
\be\label{VTTsr}
\frac{V_{TT }}{3H^2} \sim \frac{\Omega^2}{3H^2}\,,
\ee
while from \eqref{VTN1} we observe that barring cancellations, $\delta_\varphi \ll 1$ implies that
\be\label{VTN_slowroll}
   \frac{V_{TN}}{3H^2}\sim \frac{ \Omega}{H} \,, \qquad {\rm and} \qquad 
     \nu \ll 1\,.
\ee
Hence, $\nu$ is as a new slow-roll parameter in multifield inflation: the turning rate is guaranteed to be slowly varying in slow-roll.

\subsubsection*{No $\eta$-problem in mulfield inflation}

Above we have explicitly derived the conditions for long lasting, slow-roll multfield inflation, namely \eqref{phislow}-\eqref{VTN_slowroll}. So long as these conditions are satisfied, long-lasting slow-roll trajectories are guaranteed. Note in particular that this does not impose any constraint on $\eta_V$, defined in \eqref{etaV}. Indeed, the minimal eigenvalue of ${\mathbb M}$ satisfies the condition 
\be
\lambda_{\rm min} \leq U_a{\mathbb M}^a_{\,\,b}U^b\,,
\ee
where $U^a$ is an arbitrary unit vector. Taking $U^a=T^a$, the tangent to the trajectory, we have that $\lambda_{\rm min} \leq {\mathbb M}_{TT}= V_{TT}/V$. Thus, there arises the possibility of fat inflation, introduced in  \cite{Chakraborty:2019dfh}. There, it was shown that  a sufficient condition to obtain large turns  is to have the minimal eigenvalue of  ${\mathbb M}$ be much larger than one, that is $\lambda_{\rm min} \gg 1$, which thus  implies $V_{TT}/V\gg1$, which from  \eqref{VTTsr} implies $\omega\gg1$. 
Thus it is possible to have all fields heavier than the Hubble scale without spoiling long-lasting slow-roll. Moreover,  one can check that this also holds when the minimal eigenvalue is negative. That is, slow-roll inflation is possible with fat tachyonic fields, $|\lambda_{\rm min}| \gg1$, and large turning rates, $\omega\gg 1$. Note that these models satisfy the dSSC \cite{Obied:2018sgi,Garg:2018reu,Ooguri:2018wrx} along the trajectory, since $\lambda_{\rm min}\leq -{\cal O}(1)$.  
An example of this type of inflationary attractor is  angular inflation \cite{Christodoulidis:2018qdw}. 
More generally, we see that $\eta_V$ \eqref{etaV} can be either large or small, without affecting the inflationary attractor. In particular, it is not necessary to fine-tune the masses of the inflatons to be small (in Hubble units) to ensure a successful period of multifield inflation. 

Using the slow-roll expressions for the mass matrix elements $V_{TT}$ and $V_{TN}$, we can express the eigenvalues purely in terms of $\Omega$ and $V_{NN}$. If all the eigenvalues are positive, as in the case of  fat inflation, we have that $0<\det{\mathbb M}< \Tr{\mathbb M}^2/4$. Since $\det{\mathbb M} \simeq \Omega^2(V_{NN} -9H^2)$, fat inflation requires
\be\label{fatcond}
V_{NN}> 9H^2 \,.
\ee
In the opposite case, $V_{NN}< 9H^2$,  the minimal eigenvalue will be negative\footnote{Recall that a tachyon along the inflationary trajectory does not indicate an instability. The criteria for the stability of the trajectory are discussed e.g. in \cite{Christodoulidis:2019mkj}.}.

We now shift our attention toward the dynamics of rapid-turn inflationary trajectories. We begin with a discussion on two-field trajectories before analysing rapid-turn inflation in supergravity in Section \ref{sec2}.

\subsection{Two-field inflation in field theory}\label{subsec1}

Starting with the generic multi-field Lagrangian \eqref{4Daction}, we now focus on the two-field case $\phi^a = (r, \theta)$. In anticipation of examining inflation in supergravity, we refer to these fields as the {\em saxion} and {\em axion} fields, respectively. For a broad class of highly symmetric field space geometries, the kinetic term can be written as
\footnote{Note that this metric can be written in several other equivalent forms by suitable redefinition  of the  field $r$: 
\bea\label{L2}
{\mathcal L}_{\phi} & \supset& -\frac{1}{2} \left[(\partial R)^2 + f^2(R)(\partial \theta)^2\right]\, \\
& \supset&  -\frac{f^2(\rho)}{2} \left[(\partial \rho)^2 +\rho^2(\partial \theta)^2\right]\,.\label{L3}
\eea
Hence, we focus on \eqref{L1} and transform to either \eqref{L2} or \eqref{L3} by a simple redefinition of the {\em radial} coordinate.
Note further that in some cases \cite{Christodoulidis:2018qdw}, the scalar metric allegedly depends on both scalar fields. However, it only depends on a single combination of them:
\be
g_{ab} = f(\phi,\chi){\rm diag}(1,1) = \frac{1}{1-(\phi^2+\chi^2)}{\rm diag}(1,1)\,.
\ee
Therefore, it is possible to change coordinates to $\phi=r \sin\theta$ and $\chi = r \cos \theta$, such that the metric takes the form of \eqref{L3}, which in turn is equivalent to \eqref{L1}.}
\be\label{L1}
{\mathcal L}_{\phi} \supset -\frac{f^2(r)}{2} \left[(\partial r)^2 + (\partial \theta)^2\right]\,.
\ee
Note that the field space metric is independent of the coordinate $\theta$, indicating an isometry direction. The equations of motion \eqref{H} and \eqref{phis} in terms of $r$ and $\theta$ take the form
\bea\label{saxioneq}
 r'' +  \left(3-\frac{\varphi'^2}{2\Mp^2}\right)r' - \frac{f_r}{f}\left(\theta'^2-r'^2\right) +  \frac{V_{r}}{H^2 f^2} &=&0  \,,\\
  \theta'' + \left(3-\frac{\varphi'^2}{2\Mp^2}\right)\theta' +2 \frac{f_r}{f} \theta' r' +  \frac{V_{\theta}}{H^2 f^2}&  =&0  \,,
\label{axioneq}
\eea 
where primes denote e-fold derivatives, and \eqref{varphi} reduces to $\varphi'^2 = f^2(r)(r'^2 + \theta'^2)$.

\subsubsection*{Inflationary Solutions }

We now consider the possible inflationary solutions to \eqref{saxioneq} and \eqref{axioneq}.
\begin{enumerate}
\item {\em Saxion inflation}. Single field saxion inflation can occur for $\theta'\simeq 0$, with $\theta$ fixed at $\theta_0$ such that $V_\theta(r, \theta_{0})=0 \quad \forall \quad r$. In this case \eqref{axioneq} is automatically satisfied, while \eqref{saxioneq} admits slow-roll solutions given a suitable potential. An example in supergravity is discussed in \cite{KLR,RSZ}.

\item {\em Axion inflation}. A more interesting possibility occurs for solutions with $r'\simeq 0$ \footnote{ While $\theta=\theta_0$ \,(i.e. $\theta'= 0$) is always a geodesic, $r=r_0 \, (r'=0)$  is only a geodesic when $\left(f_r/f \right) \large|_{r_0}=0$. }. In this case, imposing slow-roll in $\theta$ on the equations of motion yields:
\bea
- \frac{f_r}{f} \theta'^2 +  \frac{V_{r}}{H^2 f^2} &=&0 \label{req2} \,,\\
 \left(3-\epsilon\right)\theta'  +  \frac{V_{\theta}}{H^2 f^2}  &=&0 
\label{thetaeq2} \,.
\eea
These two equations give independent constraints on $\theta^\prime$, which must be equal along the inflationary trajectory. Demanding them be equal gives the consistency condition: 
\begin{align}
    (\theta^\prime)^2 = \frac{V_\theta^2}{(3-\epsilon)^2 H^4 f^4} &= \frac{V_r}{H^2 f f_r} \,,\\
    \Rightarrow \frac{V_\theta^2}{(3-\epsilon)^2 H^2 V_r } &= \frac{f^3}{f_r}\,. \label{eq:thetaprimeconsistency}
\end{align}
This relationship vastly restricts the regions of field and parameter space satisfying slow-roll, and can be used to identify the inflationary trajectory.

We can also have different scenarios depending on the initial conditions for the saxion:
\begin{enumerate}
\item {\em Single field axion inflation.} Single field axion inflation occurs when the saxion is fixed to $r_0$ such that $V_r(r_{0}, \theta)=0$ and $(f_r/f)|_{r_{0}} =0$ so as to satisfy \eqref{req2}. An example of this solution can be obtained in  the minimal sidetrack models of \cite{Garcia-Saenz:2018ifx}.

\item {\em Multifield axion inflation.} Multifield inflation will occur whenever \eqref{req2} is satisfied with both terms non-vanishing. This can happen in two scenarios: either $V_r(r_{0}, \theta)\ne0$ and  $(f_r/f)|_{r_{0}} \ne 0$, or $V_r(r_{0}, \theta)=0$ and $(f_r/f)|_{r_{0}} =0$ with $r = r_{shift} \neq r_0$. The minimal sidetrack models of \cite{Garcia-Saenz:2018ifx} are an example of this case. In this situation, one can in principle compute  the value of $r_{shift}$ during the multifield evolution.  Note that in this case the axion follows slow-roll inflation assisted by the saxion, which can give rise to large turning rates.
These are the type of solutions we examine in what follows. 

\end{enumerate}

\item  {\em Multifield   inflation.} By appropriately choosing the initial conditions and potential, it is possible to have {\em double inflation} without substantial turning (see e.g. Appendix B of \cite{Chakraborty:2019dfh} and \cite{Ketov:2021fww} for an example in supergravity) or genuinely multifield evolution where both axion and saxion act as inflatons with interesting phenomenology.

\end{enumerate}

We are interested in {\em multifield axion inflation} models. Note that it is not possible to have a similar ``assisted saxion" multifield inflation. Note also that all two-field rapid-turn models described by the actions \eqref{L1}, \eqref{L2}, and \eqref{L3} exhibit multifield axion inflation (e.g. supergravity inspired angular inflation \cite{Christodoulidis:2018qdw}, orbital inflation \cite{Achucarro:2019mea}, hyperinflation \cite{Brown:2017osf}, sidetrack inflation \cite{Garcia-Saenz:2018ifx}).

\subsection{Rapid-turn, multifield axion inflation}

We now examine solutions to the equations for multifield axion inflation in greater detail. In this case, the inflationary trajectory is mostly aligned with the axion direction, while the saxion stays approximately constant and constitutes the direction normal to the trajectory.
Using the {\em kinematic basis} description in Section \ref{sec1}, we can express the trajectory's unit tangent and normal vectors as:
\be
T^a = \frac{1}{\varphi'}(r',\theta') \,,\qquad 
N^a = \frac{1}{\varphi'}(\theta',-r') \,. 
\ee
Since $r'\sim 0$ in multifield axion inflation, the field velocity simplifies as $\varphi'\sim f\theta'$, which reduces the unit vectors to $T^a \sim (0, 1/f)$ and $N^a \sim (1/f, 0)$. The slow-roll parameter $\epsilon_T$ \eqref{epT} is then
\be\label{etheta}
 \epsilon_T\sim 
 \frac{M_{Pl}^2}{2}\frac{1}{f^2}\left(\frac{V_{,\theta}}{V}\right)^2 \ll 1\,. 
\ee
Moreover, $\eta$ \eqref{etaH} can be written as 
\be
\eta \simeq -2\eta_T + 4 \epsilon_T\,,
\ee
where we defined 
\be\label{etaT}
\eta_T \equiv \frac{\Mp^2}{f^2}\frac{V_{,\theta\theta}}{V }\,,
\ee
which should be small during inflation $\eta_T\ll 1$. Furthermore, using the definition of $\Omega$ in \eqref{Omega}, the unit normal vector, and \eqref{thetaeq2}, one can show that \cite{Garcia-Saenz:2018ifx}
\be\label{Osimp1}
\frac{ \Omega}{H} \simeq \frac{V_r}{H^2 f^2 \theta'}\simeq -\frac{V_r}{V_\theta} \left(3-\epsilon\right)\,.
\ee
Using equation \eqref{eq:thetaprimeconsistency}, we can further write:
\be\label{Osimp2}
\frac{\Omega}{H}\simeq- \frac{V_\theta}{H^2}\frac{f_r}{f^3}\frac{1}{(3-\epsilon)}
\simeq - \Mp\sqrt{2\epsilon_T}\,\frac{f_r}{f^2}\,.
\ee
This expression for $\Omega/H$ explicitly shows that a change in the field space metric, $f(r)$, directly affects the turning rate. We will later make use of this when showing how to generate large turning rates in supergravity.

Lastly, we note that the normal projection of the potential's covariant Hessian matrix simplifies to 
\be\label{VNN}
V_{NN} = \frac{1}{f^2 }\left(V_{,rr}- V_r \frac{f_r}{f}\right) \,,
\ee
emphasising that positive eigenvalues require $V_{NN}>9H^2$, while negative ones $V_{NN}<9H^2$.

\section{Large turning rates in supergravity}\label{sec2}

We are now ready to investigate the viability of rapid-turn inflation in supergravity. Our starting point was the observation that numerical scans of supergravity models in the literature failed to find trajectories with  large turning rates. The results of the scans of supergravity models are provided in Appendix \ref{sec:survey}. In this section, we show how to understand the small turning rates of most models and provide a method to increase the turning rate. As we  discuss, these rapid-turn examples consistently have a fat tachyonic direction, ($|\lambda_{\rm min}|>1$) along the inflationary trajectory\footnote{Recall that this does not indicate an instability, similar to concave potential inflation, such as Starobinsky inflation \cite{Starobinsky:1980te}.}.
We  conjecture that rapid-turn inflationary trajectories in supergravity always occur in the presence of tachyonic directions, potentially  satisfying the dSSC.

\subsection{Multifield inflation in supergravity}

Our starting point is the supergravity Lagrangian:
\be\label{sugraS}
S= \int{d^4x \sqrt{-\tt g} \left[\Mp^2 \frac{R}{2}  - K_{i\bar \jmath} \partial_\mu\Phi^i \partial^\mu\bar\Phi^{\bar \jmath} - V(\Phi^k,\bar\Phi^k)\right]} \,,
\ee
where $K_{i\bar \jmath}$ is the K\"ahler metric and $W$ is the superpotential. The scalar potential is given in terms of the K\"ahler potential $K$ and the superpotential as
\be\label{Vsugra}
V= e^{K/\Mp^2}(K^{i\bar \jmath}D_i W D_{\bar{\jmath}} \bar{W} - 3 \vert W \vert^2 \Mp^{-2})\,,
\ee
where $D_iW= W_i + K_i/\Mp^2 W$. Here $\Phi^i$ are complex fields, of which there are generically many. To study inflation we consider an FLRW 4D spacetime, in which the equations of motion for the scalars take the form:
\bea
\ddot \Phi^i +3H \dot \Phi^i +\Gamma_{jk}^{i}\dot \Phi^j\dot \Phi^k + K^{i\bar\jmath} V_{\bar\jmath} =0 \,, 
\eea
with an additional equation of motion for the conjugate field $\bar\Phi^{\bar i}$. Here the Christoffel symbols are computed from the K\"ahler metric $K_{i\bar\jmath}$, with only $\Gamma_{ik}^{i}$ and $\Gamma_{\bar \imath\bar k}^{\bar\jmath}$ non-zero. 

The simplest setup involves a single superfield comprised of two real scalars, so our previous discussion on two-field inflation immediately applies. In this case, known as {\em sgoldstino inflation} \cite{AlvarezGaume:2011xv,Achucarro:2012hg}, the inflaton and the sgoldstino are aligned. However, inflation is generally difficult to realize with a single superfield \cite{RSZ}. In Appendix \ref{sec:sGI} we explore the possibility of sgoldstino inflation in a simple, analytically solvable model.

The next possibility involves two superfields, in such a way that during inflation, only two of the real fields evolve.
In  \cite{KLR} an interesting strategy to realise single field inflation with any potential along the direction orthogonal to the sgoldstino \cite{RSZ} was introduced. The model introduces two superfields, which act as the sgoldstino and inflaton respectively. It was shown that the three additional scalars can always be stabilised by introducing a suitable K\"ahler potential. Moreover, in \cite{Ferrara:2014kva} the sgoldstino was eliminated by introducing a nilpotent condition to the goldstino superfield. 
Note that in principle one could combine  the real fields from the different superfields to drive two-field inflation. However, such a configuration will not give rise to the type of attractors in the previous section. 
Consequently, we consider the class of orthogonal inflation models throughout this section. 

\subsection{Rapid-turn attractor in supergravity} 

We saw in Section \ref{subsec1} that slow-roll in the $r'\sim 0$ attractor implies $\epsilon_T\ll 1$, which can be written in terms of $V_\theta$ as in \eqref{etheta}. In supergravity, this is expressed in terms of the complex fields $\Phi$ and $\bar{\Phi}$ as
\be\label{eTcomplex}
\epsilon_T  \simeq  \frac{\Mp^2}{2}\frac{1}{2K_{\Phi\bar\Phi}} \left(\frac{i(V_\Phi - V_{\bar \Phi})}{V}\right)^2, 
\ee 
where $V$ is the supergravity scalar potential. The $\eta_T$ parameter \eqref{etaT} can  be written as
\be\label{etaTcomplex}
\eta_T \simeq  \frac{\Mp^2}{2K_{\Phi\bar\Phi}} \frac{(2V_{\Phi\bar\Phi} - V_{\Phi\Phi} -V_{\bar\Phi\bar\Phi})}{V} \,.
\ee 
Transforming the expression for $\omega$ in \eqref{Osimp2} to complex coordinates and neglecting factors of $\epsilon$, we obtain:
\be\label{Ocomplex2}
\frac{\Omega}{H} \simeq - \Mp^2\frac{i(V_\Phi-V_{\bar\Phi})}{V} 
\frac{\left(K_{\Phi\bar\Phi,\Phi} + K_{\Phi\bar\Phi,\bar\Phi} \right)}{(2K_{\Phi\bar\Phi})^{2}} =- \Mp \sqrt{2 \epsilon_T}\, \frac{\left(K_{\Phi\bar\Phi,\Phi} + K_{\Phi\bar\Phi,\bar\Phi} \right)}{(2K_{\Phi\bar\Phi})^{3/2}}\,.
\ee  
From this equation it is evident that large turning rates can be adjusted by tuning the K\"ahler potential. Meanwhile, the superpotential can be tuned to ensure slow-roll, i.e. $\epsilon_T\ll1$. We will see this in concrete examples below. 

In what follows, we consider large turning rates and fat inflatons in specific supergravity models. As we have identified above, tuning the K\"ahler potential and superpotential suitably can in principle generate strongly non-geodesic inflationary trajectories. In all these examples, we find  fat tachyonic  fields  along the trajectory.  

\subsection{Orthogonal inflation}

As described previously, we consider two ``orthogonal" chiral superfields, the goldstino and inflaton superfieds $S$ and $\Phi$, respectively. We denote the scalar components of these superfields with the same letter.
We can now to eliminate $S$ by either introducing a suitable K\"ahler potential to stabilise it to $S=0$, or by introducing a nilpotent condition $S^2=0$ \cite{KLR,Ferrara:2014kva}.

Consider a general K\"ahler potential and the superpotential of the form
\be
K(\Phi,\bar\Phi; S, \bar S)\,, \qquad W = S F(\Phi)\,,
\label{eq:superpotentialAssumption}
\ee
where the K\"ahler potential is separately invariant under the transformations $S\to -S$. 
This ensures that the K\"ahler potential is a function of $S^2+\bar S^2$ and $S\bar S$ (if $S$ is nilpotent, $K$ depends only on $S\bar S$); for now we do not make further assumptions on $\Phi$.
In this case, $K_S = K_{\bar S} =0$ at  $S=0$, and derivatives of the superpotential reduce to:
\be
D_SW = F(\Phi) \,, \qquad D_\Phi W = 0\,,
\ee
The scalar potential then takes the simple form:
\be\label{Vsugra1}
V= e^{K(\Phi,\bar\Phi,0,0)/\Mp^2} K_{S\bar S}^{-1}(\Phi,\bar\Phi,0,0)\, |F(\Phi)|^2\,.
\ee
Assuming that the K\"ahler potential is shift symmetric in $\Phi$, such that it is a function of $(\Phi+\bar\Phi)$ only, we have the simplifications $K_\Phi=K_{\bar\Phi}$, $K^{S\bar{S}}_{\,\,\,\,\,\,\,\,\,,\Phi} =K^{S\bar{S}}_{\,\,\,\,\,\,\,\,\,,\bar\Phi}$, etc. The expressions for $\epsilon_T$ and $\omega$ in \eqref{eTcomplex} and \eqref{Ocomplex2} reduce to:
\be\label{epsthetaphi2}
\epsilon_T = -\frac{M_{Pl}^2}{4K_{\Phi\bar\Phi} }\left(\frac{ F_{\Phi}\bar{F}-F\bar{F}_{\bar\Phi} }{F\bar F}\right)^2\,,
\ee
\bea\label{Ocomplex4}
\frac{\Omega}{H} 
&\simeq&- \Mp^2\frac{i\left(  F_\Phi\bar{F}-F\bar{F}_{\bar\Phi}    \right)  }{F\bar F}
\frac{\left(2K_{\Phi\bar\Phi,\Phi} \right)}{(2K_{\Phi\bar\Phi})^{2}}
\simeq  -\Mp\sqrt{2\,\epsilon_T} \, \frac{\left(2K_{\Phi\bar\Phi,\Phi} \right)}{(2K_{\Phi\bar\Phi})^{3/2}} \,.
\eea
From here we again observe that slow-roll is attainable by suitably tuning the superpotential, while large turning rates can be obtained by tuning the K\"ahler potential.
One can also write the eigenvalues  in terms of derivatives of $K$ and $W$, making use of the slow-roll conditions. However, it is not simple to understand analytically why there is always a fat  tachyonic direction. 

\subsection{Generating large turning rates in supergravity}\label{sec:larget}

We now discuss two models where we demonstrate how to use our discussion above to generate large turning rates. As we will see, these  models are stable, while they always feature a fat tachyonic Hessian element along the inflationary trajectory.

\subsubsection{No-scale inspired model}
\label{sec:noScaleModels}

Let us consider \eqref{eq:superpotentialAssumption} with the following K\"ahler potential:
\be\label{Knoscaletoy}
 K = -\,3\, \alpha  \Mp^2\log[(\Phi+\bar{\Phi})/\Mp] + S \bar{S}\,,
\ee
which corresponds to no-scale supergravity for $\alpha=1$ \cite{Cremmer:1983bf}. For a general $\alpha>0$, the field space  curvature is given by ${\mathcal R}=-4/(3\alpha)$. The potential \eqref{Vsugra1} is
\be
V= \frac{\Mp^{3\alpha}\,|F|^2}{(\Phi+\bar\Phi)^{3\alpha}}\,,
\ee
while the turning rate \eqref{Ocomplex4} is
\be\label{omegaEx1}
\frac{\Omega}{H} \simeq  \frac{2\sqrt{\epsilon_T}}{\sqrt{3\,\alpha }} \,.
\ee
As anticipated, choosing an appropriate K\"ahler potential allows us to generate large turning rates. 
This requires a sufficiently small $\alpha\ll1$, which consequently yields a large negative curvature.
We have checked that this is the case for a wide variety of superpotentials $F(\Phi)$. For clarity we now concentrate on the simple choice\footnote{The exact form of $F(\Phi)$ is unimportant for supporting inflation, as can be seen in several of the families of models in Table \ref{tab:SUGRAscan}.}: 
\be
F(\Phi) = p_0+ p_1 \Phi\,.
\ee
%%%%%
In terms of real fields $\Phi = r+i\theta$,  the scalar potential and field space metric are given by
\be\label{noS}
    V = \Mp^{3\alpha} \, \frac{\left[p_1^2 \theta^2 + (p_0 + p_1 r)^2\right]}{8^\alpha r^{3\alpha}},   \qquad 
    g_{ab} = \frac{3\,\alpha\,\Mp^2}{2 r^2} \delta_{ab}\,.
\ee
Before examining inflationary solutions, we first consider the masses of the inflatons. As seen in the previous section, for the attractor with $r'\sim0$, $V_{NN}$ can be written as in eq.~\eqref{VNN}. In the small-$\alpha$ limit this simplifies to
\begin{align}\label{VNNs}
    \frac{V_{NN}}{H^2} - 9 \simeq \frac{ 3\,r \omega}{\theta}\left( \sqrt{2} - \frac{3}{\epsilon}\right) - 9 + \mathcal{O}(\alpha).
\end{align}
The sign of the expression above \eqref{VNNs} determines the sign of the determinant, and the number of positive eigenvalues.
When $\epsilon$ is small as required for inflation, \eqref{VNNs} is manifestly negative. Additionally, whenever $\epsilon$ is small,  $\omega$ is large, and $\alpha$ is small, we find that $V_{NN} < 9H^2$. This fixes the Hessian's eigenvalues to have opposite signs, implying the existence of a tachyonic direction, which, as we will see, is large in Hubble units.

%%%%%%%%%%%%%%%%%%%%%%%%%%%%%%%%%%%%
%%%%%%%%"No-scale" rapid turns model alpha plot %%%%%%%%%%%
%%%%%%%%%%%%%%%%%%%%%%%%%%%%%%%%%%%%
\begin{figure}[H]
  \begin{center}
    \includegraphics[width=0.5\linewidth]{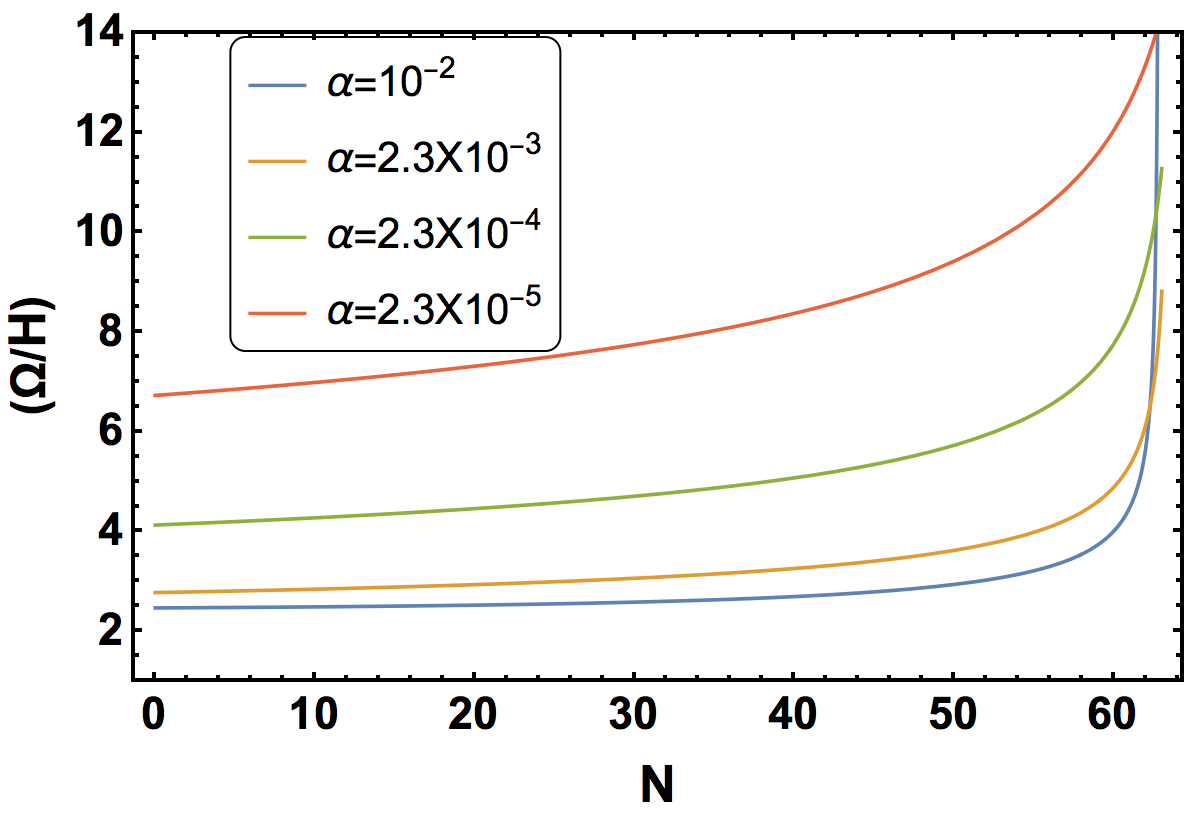}  
 \end{center}
\caption{Here we show  the dimensionless turning rate $\Omega/H$ for different values of $\alpha$ for the no-scale inspired model \eqref{noS}. In all these cases, inflation lasts at least 60 e-folds and we plot the turning rate in the last 60 e-folds. For $\alpha\gtrsim10^{-2}$, inflation lasts less than 60-folds for the same values of the  parameters ($p_0, p_1$). } \label{NoScalealpha}
\end{figure}
%%%%%%%%%%%%%%%%%%%%%%%%%%%%%%%%%%%%%%%%
%%%%%%%%%%%%%%%%%%%%%%%%%%%%%%%%%%%%
\begin{figure}[H]
  \begin{center}
    \includegraphics[width=0.55\linewidth]{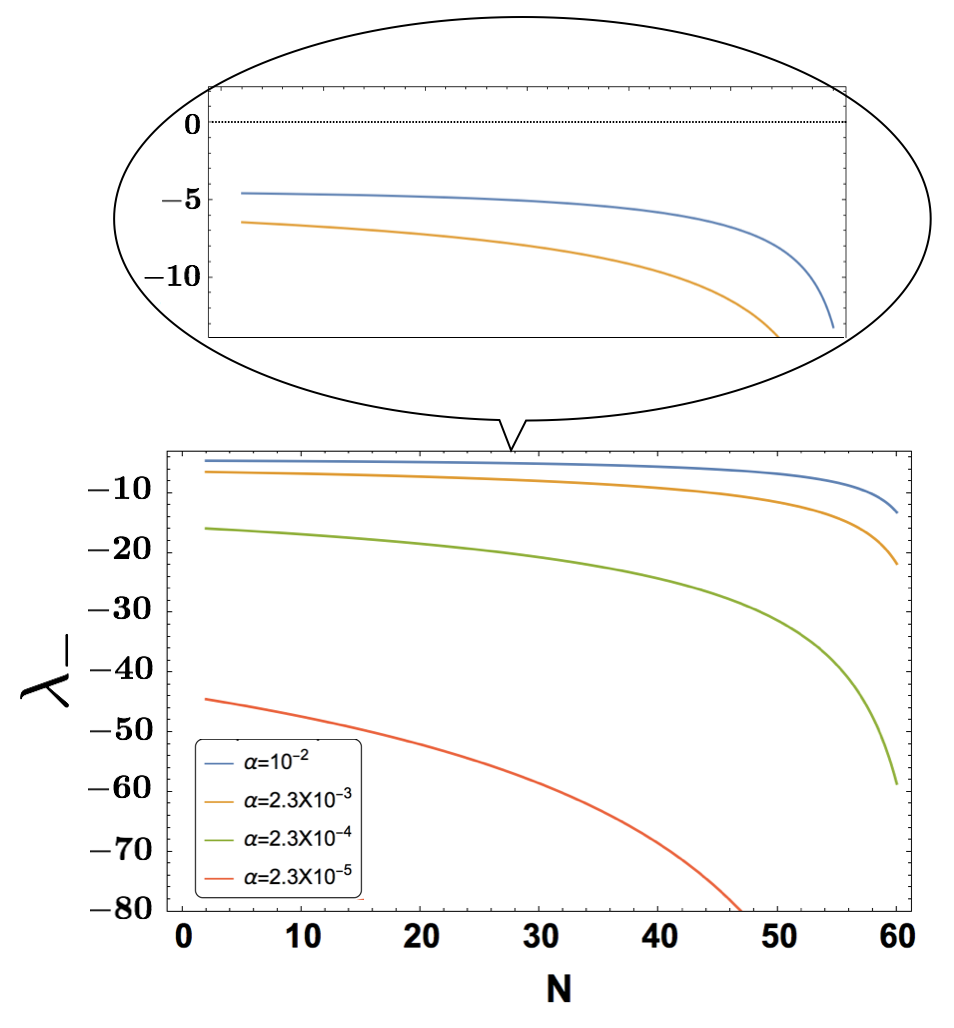}  
 \end{center}
\caption{We show here the minimal eigenvalue for different values of $\alpha$ as in figure \ref{NoScalealpha}. It is clear that they are always fat and tachyonic and increase (in absolute value) as the turning rate increases.  Thus these models satisfy the dSSC along the inflationary trajectory. } \label{NoScaleEigen}
\end{figure}
%%%%%%%%%%%%%%%%%%%%%%%%%%%%%%%%%%%%%%%%
%%%%%%%%%%%%%%%%%%%%%%%%%%%%%%%%%%%%]

In Figure \ref{NoScalealpha}, we show the turning rate for different values of $\alpha$. For all the values shown, inflation lasts at least 60 e-folds; we plot the turning rate in the last 60 e-folds. For values of $\alpha\gtrsim 10^{-2}$, inflation lasts less than 60 e-folds for the same values of the parameters  $(p_0, p_1)$. As discussed previously, it is possible to generate strongly non-geodesic trajectories by tuning $\alpha$, which changes the field space curvature, ${\cal R}=-4/3\alpha$.  On the other hand, one of the masses is always tachyonic and large along the inflationary trajectory. We show in Figure \ref{NoScaleEigen} the minimal eigenvalue of \eqref{Hessian}, which clearly satisfies the dSSC.

\subsubsection{The EGNO model}
\label{secEGNO}

We now discuss the only supergravity model we are aware of with a dimensionless turning rate larger than one: the EGNO model of \cite{EGNO}.  The original Lagrangian obeys the symmetries of  $K(S, \bar S)$ discussed previously, though in principle there is not necessarily a shift symmetry in $\Phi$. However, the parameters that yield turning rates larger than one and long-lasting inflation do admit a shift symmetry. In this region of parameter space, we can make use of the expressions found in our above analysis of rapid-turn inflation in supergravity.

\subsection*{The Lagrangian}
Setting $\Mp=1$ in this subsection for simplicity, the K\"ahler potential and superpotential for the EGNO model are
\bea\label{Kegno}
\hskip-1cm K&=& -3\,\alpha\,\log{\left[\Phi +\bar \Phi - c \left[(\Phi +\bar \Phi - 1) \cos{(p)}- i (\Phi -\bar \Phi)\sin{(p)} \right]^4\right]} \!+\! \frac{S\bar S}{(\Phi +\bar \Phi)^{3}} \,,%  
\\ %\\
\hskip-1cmW &=& S F(\Phi) \,, \qquad F(\Phi)= \sqrt{\frac{3}{4}} \frac{M}{a} (\Phi-a )\,.\label{Wegno}
\eea
Following our previous discussion, we  introduced the parameter $\alpha$, which allows for tuning to obtain large turning rates. The parameters $p$, $c$, $a$, and $M$ are arbitrary constants.

For $p=0$, the K\"ahler potential and superpotential satisfy the symmetries discussed previously\footnote{In Appendix \ref{sec:survey} we scan over $p$ as a free parameter.}. The scalar potential is given by \eqref{Vsugra1}: %
\bea
V&=& \frac{3}{4}\frac{M^2}{a^2}\frac{(\Phi+\bar\Phi)^{3}\,(a-\Phi)(a-\bar\Phi)}{\left(\Phi +\bar \Phi - c \left[(\Phi +\bar \Phi - 1)  \right]^4\right)^{3\alpha}} 
 \nonumber  \\
&=& \frac{6M^2 r^3 \left(2r-c(1-2r)^4\right)^{-3\alpha} \left((a-r)^2 + \theta^2\right)    }{a^2}\,,
\eea
where the superfield $\Phi$ is expanded as $\Phi= r +i \theta$. We observe that the potential has a minimum at $(r_{\rm min}, \theta_{\rm min}) =(a, 0)$ for any value of $\alpha$.

This model has a non-trivial field space curvature, ${\cal R}$, along $S=\bar S=0$, which depends on the value of  $\alpha$ and $c$. When $c=0$ we have ${\mathcal{R} }=-4/3\alpha$, while for $c\to\pm\infty$ the curvature is ${\cal{R}} =-1/3\alpha$. Additionally, ${\cal{R}} \to -1/3\alpha$ as $r \to \infty$. Interestingly, the curvature can be very large and positive or negative depending on the values of $c$ and $\alpha$. In particular, the $\alpha=1$ inflationary trajectory in  \cite{EGNO} has ${\cal R}>0$ (see Figure \ref{EgnoRs}). Note that although the curvature changes sign, the metric is always positive. 

The EGNO model has $\alpha=1$, $a=1/2$, $c=1000$, and $p=0$, which sets the dimensionless turning rate at $\omega\simeq 1.5$ (see the left panel of Figure \ref{EgnoO}). For $p\ne0$, our scan in 
Appendix \ref{sec:survey} found smaller turning rates whenever $p$ is far from a multiple  of $\pi$ (see Fig.~\ref{omega_p}).

\begin{figure}[h]
  \begin{center}
    \includegraphics[width=0.6\linewidth]{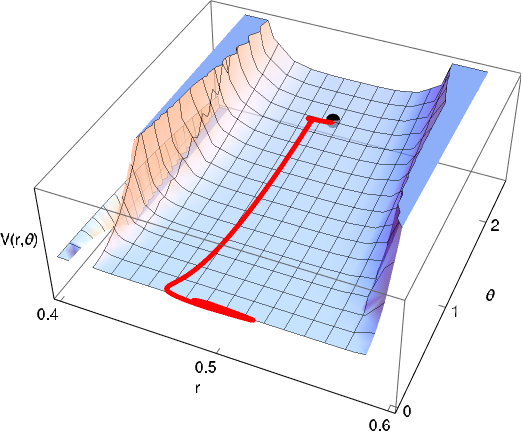} \qquad 
 \end{center}
\caption{ EGNO potential and inflationary trajectory for the parameters $\alpha=1$, $a=1/2$, $M=10^{-3}$, $c=1000$, and $p=0$ as in the original model \cite{EGNO}. The initial conditions are $r_{\rm ini} = a$, $\theta_{\rm ini} = 5 a \sqrt{2/3} $, yielding $N_{tot}=87$. 
}\label{EgnoV}
\end{figure}

\begin{figure}[h]
  \begin{center}
    \includegraphics[width=0.48\linewidth]{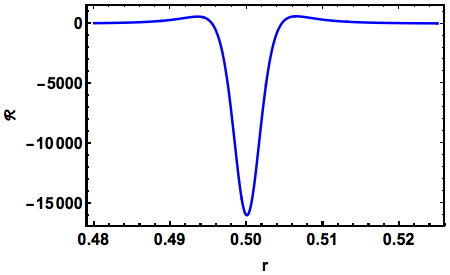} \qquad 
 %  %\begin{center}
      \includegraphics[width=0.46\linewidth]{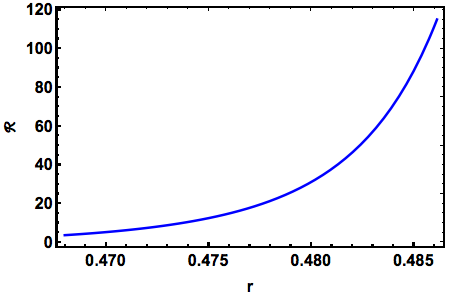} %
 \end{center}
\caption{ Curvature around the inflationary region (right) and during the last 60-efolds of inflation (left) in the EGNO model for the parameter values given in Figure \ref{EgnoV}.}\label{EgnoRs}
\end{figure}

\begin{figure}[h]
  \begin{center}
    \includegraphics[width=0.48\linewidth]{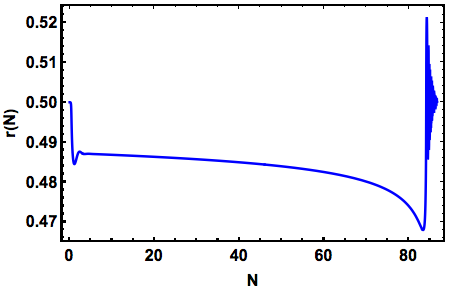} \qquad 
 %  %\begin{center}
      \includegraphics[width=0.46\linewidth]{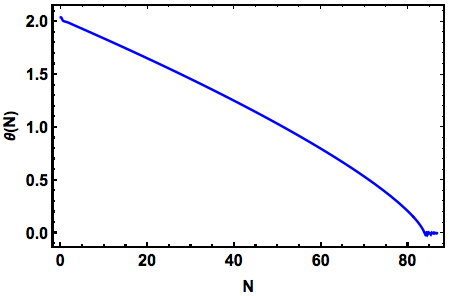} %
 \end{center}
\caption{ Trajectories for $r$ and $\theta$ in the original EGNO model for the parameters and initial conditions given in Figure \ref{EgnoV}. }\label{EgnoTs}
\end{figure}

We emphasize that $\omega$ can be increased or decreased by tuning the K\"ahler potential as in \eqref{Ocomplex4}. Since it depends on $c$, we may increase or decrease $\omega$ by increasing or decreasing $c$. For example, when $c=10$, the turning rate drops below one, $\omega\lesssim1$.  In Figure \ref{EgnoV} we show the potential and inflationary trajectory for the example in \cite{EGNO} with $\alpha=1$, $a=1/2$, $M=10^{-3}$, $c=1000$, $p=0$, and initial conditions as indicated in the figure. The evolution of the fields $r$ and $\theta$ for this example is shown in Figure \ref{EgnoTs}.

%%%%%%%%%%%%%%%%%%%%%%%%%%%%%%%%%%%%
%%%%%%%%%%%Egno Potential plot and trajectory %%%%%%%%%%%
%%%%%%%%%%%%%%%%%%%%%%%%%%%%%%%%%%%%
%%%%%%%%%%%%%%%%%%%%%%%%%%%%%%%%%%%%%%%%
%%%%%%%%%%%%%%%%%%%%%%%%%%%%%%%%%%%%

%%%%%%%%%%%%%%%%%%%%%%%%%%%%%%%%%%%%
%%%%%%%%%%%%%Cu rvature plots egno original %%%%%%%%%%%
%%%%%%%%%%%%%%%%%%%%%%%%%%%%%%%%%%%%
%%%%%%%%%%%%%%%%%%%%%%%%%%%%%%%%%%%%%%%%
%%%%%%%%%%%%%%%%%%%%%%%%%%%%%%%%%%%%

%%%%%%%%%%%%%%%%%%%%%%%%%%%%%%%%%%%%
%%%%%%%%%%% trajectory plots egno original %%%%%%%%%%%
%%%%%%%%%%%%%%%%%%%%%%%%%%%%%%%%%%%%
%%%%%%%%%%%%%%%%%%%%%%%%%%%%%%%%%%%%%%%%
%%%%%%%%%%%%%%%%%%%%%%%%%%%%%%%%%%%%

\begin{figure}[h]
  \centering
   \includegraphics[width=0.99\linewidth]{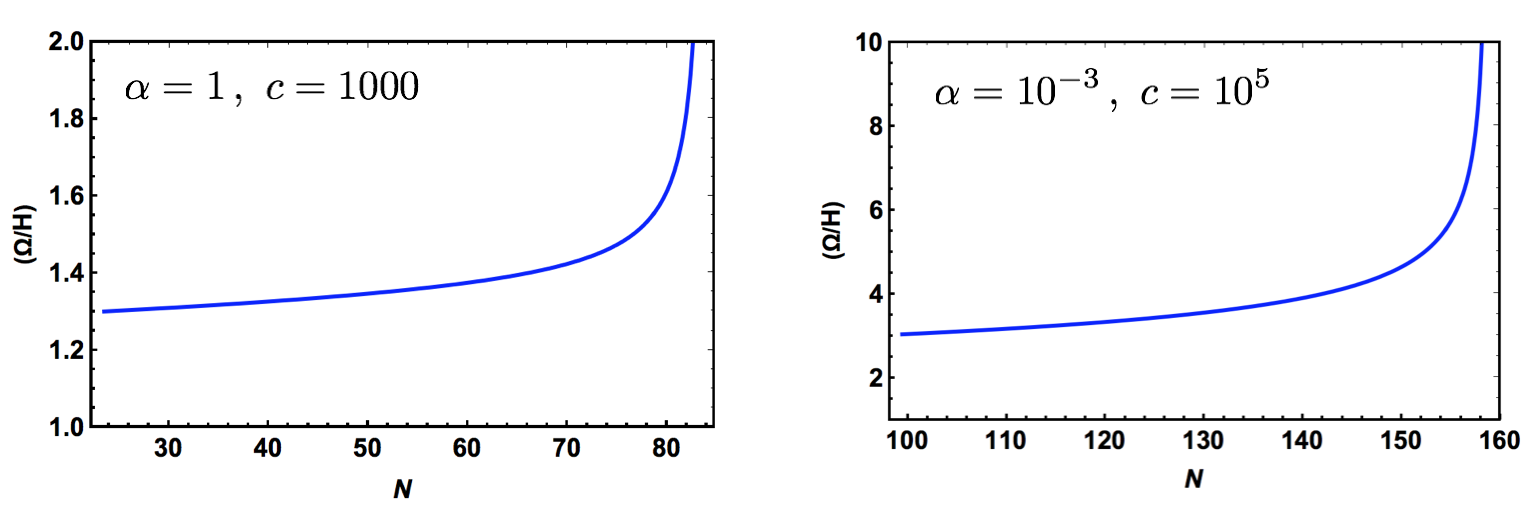}
\caption{Turning rate  in the original EGNO model for two sets of parameters. On the left, we use the parameters and initial conditions given in Figure \ref{EgnoV} and find $\omega(N_{end})\simeq 3$, where $N_{end}$ is the end of inflation, i.e. $\epsilon=1$. On the right, we use $\alpha=10^{-3}$, $c = 10^{5}$ to increase the turn rate up to $\omega(N_{end})\simeq 42$.}\label{EgnoO}
\end{figure}

%%%%%%%%%%%%%%%%%%%%%%%%%%%%%%%%%%%%%%%%
%%%%%%%%%%%%%%%%%%%%%%%%%%%%%%%%%%%%

\begin{figure}[h]
  \centering
   \includegraphics[width=0.99\linewidth]{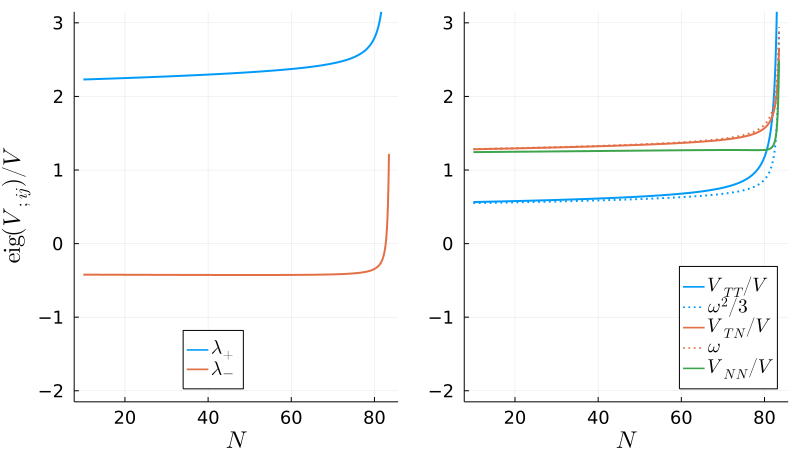}
\caption{On the left, we show the Hessian eigenvalues on the trajectory of the EGNO model for the parameters and initial conditions of Figure \ref{EgnoV}.  On the right, for the same initial conditions, we show the kinematic-basis Hessian elements during the evolution and compare them to their dynamical equivalents when available (i.e., Eqs. \eqref{VTTsr} and \eqref{VTN_slowroll}).}
\label{fig:EgnoMasses}
\end{figure}

  %%%%%%%%%%%%%%%%%%%%%%%%%%%%%%%%%%%%%%%%
%%%%%%%%%%%%%%%%%%%%%%%%%%%%%%%%%%%%

It is now evident that in the ENGO model, we can turn on $\alpha$ to  generate rapidly turning trajectories. As a concrete example, in the right panel of Figure \ref{EgnoO} we show the value of $\omega$ for a smaller value of $\alpha$ and a larger value of $c$. The field space curvature increases for these values, but as in the original model the metric is always positive.
Moreover, in the original EGNO model, the minimal eigenvalue is negative and of order $\lambda_- \sim -0.4$ (see Figure \ref{fig:EgnoMasses}), while for the modified EGNO model  with larger turning rate (left panel, Figure  \ref{EgnoO}), the smallest eigenvalue is larger in absolute value, $\lambda_- \sim -3 $  thus satisfying the dSSC along the inflationary trajectory.  Nevertheless the potential does not  satisfy it  globally. 
Away from the trajectory, the potential has neither a large gradient parameter $\epsilon_V$ nor a large tachyonic direction.

\FloatBarrier
\section{Conclusions}\label{sec3}
Strongly non-geodesic inflationary trajectories in multifield inflation have attracted revived interest recently from theoretical and phenomenological perspectives.
However to date, rapid-turn multifield models in supergravity and string theory are scarce. On the supergravity side, the only model we are aware of with an order one turning rate $\omega\gtrsim 1$ is the EGNO model \cite{EGNO} that we discussed in Section \ref{secEGNO}.  On the string theory side, the only available model is the multifield fat inflation D5-brane model introduced in \cite{Chakraborty:2019dfh}. 

In the present work we have systematically analysed rapid-turn  inflation in supergravity as a first step toward understanding multifield inflationary attractors in string theory. 
In Section \ref{sec1}, we revisit the slow-roll conditions in multifield inflation and showed that light inflatons (in Hubble units) are not required to ensure sustained inflation. That is, the  $\eta_V$ parameter \eqref{etaV} does not need to be small in multifield inflation as commonly assumed.  
We further discuss in detail the large turn inflationary attractor in effective field theory and  study the forms of two-field inflation that may occur in supergravity Lagrangians and focus on multifield axion inflation for its relation to well-known rapid-turn inflationary models in the literature.

In Section \ref{sec2} we then introduce multifield inflation in supergravity. For concreteness, we focus on a large class of two-superfield supergravity models, in which the inflaton is orthogonal to the sGoldstino direction \cite{KLR,Ferrara:2014kva,RSZ}. In this class of models, inflation occurs along a single superfield direction, i.e. along two real directions.  Using our discussion in effective field theory, we find expressions for the slow-roll parameter $\epsilon$ and dimensionless turning rate $\omega$ in terms of derivatives of the K\"ahler potential \eqref{epsthetaphi2} and the superpotential \eqref{Ocomplex4}.
From these expressions we observe that one can tune the superpotential $F(\Phi)$ to ensure a small $\epsilon$, while independently tuning of the  K\"ahler potential to increase the turning rate.

We find a large class of models with a high turn rate, a large field space curvature, and a fat tachyonic mass, that is, with $\eta_V \lesssim -1$.
This class matches all instances of rapid-turn inflation found in our survey of supergravity models in Appendix \ref{sec:survey}.
We study in detail two of these models:  a no-scale-inspired model and the EGNO model.
In both cases, we show that the turn rate increases as the field space curvature increases and that one of the masses is always tachyonic when slow-roll and rapid-turn are valid approximations. This tachyon does not destabilise the trajectory.

In both supergravity rapid turn models we discussed above, we have tuned  by hand the parameters need to get strongly non-geodesic inflationary trajectories\footnote{Non-geodesicity constraints were recently studied in \cite{Calderon-Infante:2020dhm} for trajectories that asymptote to infinity, which therefore cannot be applied to inflation where all trajectories end at the minimum of the scalar potential.}. However, these can only be considered as toy models, as such small values of $\alpha$ do not occur in theoretically motivated models of supergravity or string theory. Interestingly, tuning of the superpotential and K\"ahler potential to achieve long lasting inflation and large turns, gives rise to fat tachyonic fields. 

In the main text, we focused on a large class of supergravity models that were useful to illustrate our findings. We expect however that similar arguments apply to more general models\footnote{For example, K\"ahler inflation \cite{Conlon:2005jm} is a small turn attractor with a light tachyonic inflaton (that is, $\eta_V\ll1$)  in the two field case \cite{Bond:2006nc,Blanco-Pillado:2009dmu}. Following our discussion,  tuning the K\"ahler potential by hand, it should be possible to find a strong non-geodesic inflationary attractor.}. In Appendix \ref{sec:sGI}, we also discuss  a single superfield example where we can see that inflation with large turns cannot be achieved.

These results, together with our survey of a wide variety of supergravity models, lead us to conjecture that rapid-turn inflation is rare in theoretically motivated supergravity constructions.
This is the primary conclusion of this paper.
When allowing for large field space curvature, rapid-turn inflation becomes possible with  $\eta_V\lesssim -{\cal O}(1)$. This appears to be a ubiquitous feature of rapid-turn inflation in supergravity; tachyons are also a feature of de Sitter constructions in supergravity \cite{Andriot:2021rdy}. The models we have examined in Section \ref{sec2} do not satisfy the refined de Sitter conjecture globally as they have points with positive masses and  $\epsilon_V\sim {\cal O} (10^{-4})$; however, rapid-turn inflationary trajectories do not exist in that region.

\acknowledgments{We are grateful to Irene Valenzuela and Gianmassimo Tasinato for insightful discussions. We would also like to thank Perseas Christodoulidis for  comments on the first version of this paper. The National Science Foundation supported the work of VA, RR and SP under grant number PHY-1914679. Part of the work of SP was performed at the Aspen Center for Physics, which is supported by National Science Foundation grant PHY-1607611. RCh is supported by an STFC postgraduate scholarship in data intensive science.   IZ is partially supported by STFC, grant ST/P00055X/1.}

\begin{appendix}

\section{Single superfield model}\label{sec:sGI}

In order to understand the viability of large turning rates analytically, we now examine models consisting of a single superfield $\Phi$, i.e. multifield  sGoldstino inflation. We consider models with a K\"ahler potential given by
\begin{align}
	K = -3\alpha\log[(\Phi + \bar{\Phi})/\Mp].
\end{align}
For simplicity, we take the superpotential to be a monomial in $\Phi$,
\begin{align}
	W = M^{3-n} \Phi^n,
\end{align}
where $n$ is an integer. Expanding the superfield into real and imaginary parts, $\Phi = r + i\theta$, the resulting scalar potential is
\begin{align}
	V = 3M^{2 (3-n)} \frac{ \left(\theta ^2+r^2\right)^{n-1} }{2^{3\alpha} r^{3\alpha}} \left[ \left(\frac{(\alpha-2 n/3)^2}{\alpha}-1\right)r^2 + (\alpha-1)\theta ^2\right],
\end{align}
and the metric on real fields takes the form of \eqref{L1} with $f = \sqrt{\frac{3\alpha}{2r^2}}$.

Axion inflation solutions require the consistency of both $\theta'$ expressions in \eqref{req2} and \eqref{thetaeq2}. Denoting these expressions as $\theta'_1$ and $\theta'_2$ respectively, we find them to be
\begin{align}
	\theta'_1 = &\sqrt{\frac{2}{\alpha \left(\theta ^2+r^2\right) \left(9\alpha^2 \left(\theta ^2+r^2\right)-3\alpha \left(3 \theta ^2+(4 n+3) r^2\right)+4 n^2 r^2\right)}} \\
	&
	\begin{aligned}
		\times \bigg[ &r^2 \bigg(-27 \alpha^2 \left(\theta ^2+r^2\right) \left(\theta ^2+(2 n+1) r^2\right)+27\alpha^3 \left(\theta ^2+r^2\right)^2 \\
		& + 6 \alpha n r^2 \left(\theta ^2 (2 n+7)+(6 n+3) r^2\right)-8 n^2 r^2 \left(\theta ^2+n r^2\right)\bigg) \bigg]^{1/2}
	\end{aligned}
	\\
	\theta'_2 = &-\frac{4 \theta  n r^2 \left(9\alpha^2 \left(\theta ^2+r^2\right)-9 \alpha \theta ^2+3\alpha (1-4 n) r^2+4 (n-1) n r^2\right)}{3\alpha \left(\theta ^2+r^2\right) \left(9\alpha^2 \left(\theta ^2+r^2\right)-3\alpha \left(3 \theta ^2+(4 n+3) r^2\right)+4 n^2 r^2\right)}.
\end{align}
When these expressions are equal, slow-roll trajectories are constrained to be of the form
\begin{align}
	r = c_n \theta,
\end{align}
where $c_n$ is a proportionality constant depending on the order $n$ of the monomial superpotential and $\alpha$. Although the expressions are unwieldy for $n \geq 2$, we present the proportionality constant for the $n=1$ case below:
\begin{align}
\hskip-0.3cm 
c_1 \!= \!\pm \sqrt{3}\alpha^{3/4} \sqrt{\frac{ \!\sqrt{\alpha}\!\left(\!-81 \alpha^3+297 \alpha^2\!-\!303 \alpha\!+\!87\right) \!+\! \sqrt{3}\sqrt{(\!-405 \alpha^3+1242 \alpha^2\!-\!933 \alpha+288)(\alpha\!-\!1\!)^2} }{ (3\alpha-2)(9\alpha^2-21 \alpha+4)^2 } }.
\end{align}
These solutions are only real-valued for $1 < \alpha < \frac{7+\sqrt{33}}{6}$, which corresponds to a negative potential. Hence, we can exclude slow-roll axion inflation from models with an $n=1$ monomial superpotential. For higher order monomials with $n=2,3,4,5$, our numerical scans similarly find that the potential is either negative or complex wherever the solutions are real-valued. We therefore expect slow-roll axion inflation to be forbidden in this entire class of models with monomial superpotentials.

\section{Results from survey of supergravity models}
\label{sec:survey}

To study the possibility of rapid-turn inflation in supergravity models, we surveyed several from the literature (\cite{EGNO,Kallosh:2010ug,Kallosh:2014vja,German:2019aoj,Blumenhagen:2015qda,Aldabergenov:2020atd,Blanco-Pillado:2004aap}), as well as studied several ad-hoc models of our own creation. 
After constructing the potential and field space metric in terms of real fields, we scanned a wide region of allowed field and parameter space for each model.
This was achieved using an efficient differential-evolution optimizer in \texttt{BlackBoxOptim.jl}\footnote{\url{https://github.com/robertfeldt/BlackBoxOptim.jl}}, assuming ``good'' initial inflationary points minimized one of several cost functions.
We performed several scans with each choice of cost function, first assuming the initial velocities to follow the rapid-turn solution \cite{Bjorkmo:2019fls,Aragam:2020uqi}. This cost function can be written
\begin{align}
    \mathrm{cost}_\mathrm{rapid-turn}(\vec{\phi},\mathrm{parameters}) = \frac{\ddf_v}{3 H \df_v} + A \epsilon + B|\eta| + C|\nu| + D (\Omega/H)^{-2},
\label{eq:RTAcost}
\end{align}
where $v^a$ is the potential gradient unit vector and $\dot{\phi}_v = v^a\dot{\phi}_a$.
This expression is small when the solution admits both slow-roll and rapid-turn, and with $A,B,C,D$ chosen to weight each term's relative contribution. Typical values chosen were $A=100,B=10,C=D=1$, though small changes in these values did not affect the result. For details on our numerical method of constructing the zero-torsion rapid-turn inflationary solution at a given point in parameter space, see \cite{Aragam:2020uqi}.

As alternatives, we also examined cost functions to prefer high masses, with no inflationary considerations:
\begin{align}
\mathrm{cost}_\mathrm{fat}(\vec{\phi},\dot{\vec{\phi}},\mathrm{parameters}) = 10^{10} ( \text{\# negative eigvals of }V^a_b) + \frac{1}{|\text{min. eigval of }V^a_b|},
\label{eq:Fatcost}
\end{align}
and an empirical cost function, with no assumptions about the initial conditions other than the velocities were small enough to allow inflation to begin, i.e. the initial $\epsilon < 1$:
\begin{align}
 \mathrm{cost}_\mathrm{empirical}(\vec{\phi},\dot{\vec{\phi}},\mathrm{parameters}) = \frac{1}{N_\textrm{end}} + \frac{J}{\omega_\textrm{end}},
 \label{eq:Empiricalcost}
\end{align}
where $J = 10^6 \omega_\textrm{end}$ if the total number of e-foldings $N_\textrm{end} < 60$ and is otherwise $1$, and $\omega_\textrm{end}$ is the lowest value of $\omega$ recorded during the final 10 e-folds of evolution. Numerical integration was paused when $\epsilon=1$ or after 60 e-folds, whichever occurred first.
The empirical and high-mass scans treated the fields' initial velocities as free parameters, rather than determining them with the rapid-turn solution.

For each choice of cost function, at several hundred of its minimizing field and parameter values, we evolved the classical equations of motion using the publicly-available multi-field inflationary dynamics code \texttt{Inflation.jl} \cite{Inflationjl} and recorded the number of e-folds as well as the lowest turn rate recorded during the final 10 e-folds, $\omega_\text{end}$.
The empirical cost function proved to be the most successful at finding inflationary points with slow-turn inflation, while the rapid-turn cost function was comparably effective at finding rapid-turn inflation. The fat cost function was the least effective at finding inflationary initial conditions, suggesting the correlation between high mass points and inflationary points is not a strong one in these models.

Below we display each scanned model's K\"ahler and superpotential, its reference when available, and the best solution as ranked by the empirical cost function, even if found in a scan using one of the other cost functions. Many published models restrict the field space to the regime with $S=0$, but in these scans we have avoided that limitation when possible, rendering some of these models non-inflationary.
Because the scale of the inflationary potential does not affect the background evolution, coefficients common to all terms in the superpotential have been neglected when possible.
Although not indicated in the table, none of the found rapid-turn phases were fat inflation.
In each model, its real fields are defined from the complex fields as $\Phi = \phi^1 + i \phi^2$, $S = \phi^3 + i \phi^4$. We denote the initial e-fold velocities by $\pi^a \equiv \dot\phi^a / H$. For notational convenience we set $\Mp = 1$.

%%%%%%%%%%%%%%%%%%%%%%%%%%%%%%%%%%%%
%%%%%%%%"omega^2 as function of p, EGNO %%%%%%%%%%%
%%%%%%%%%%%%%%%%%%%%%%%%%%%%%%%%%%%%
\begin{figure}[h]
  \begin{center}
    \includegraphics[width=0.7\linewidth]{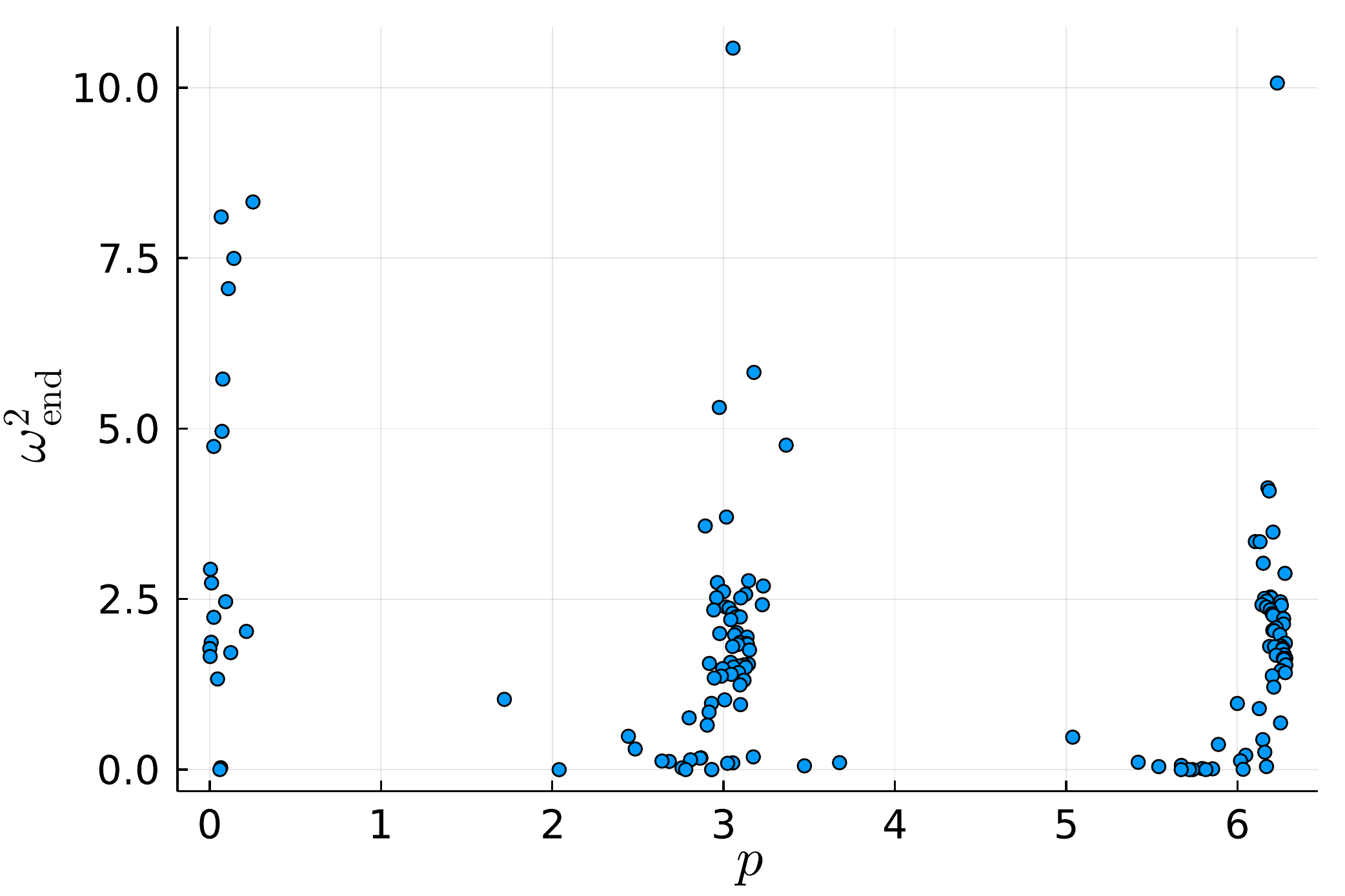}  
 \end{center}
\caption{Behaviour of $\omega_{\rm end}^2$ for different values of $p$ in the EGNO model in Table \ref{tab:SUGRAscan}. } \label{omega_p}
\end{figure}
%%%%%%%%%%%%%%%%%%%%%%%%%%%%%%%%%%%%%%%%
%%%%%%%%%%%%%%%%%%%%%%%%%%%%%%%%%%%%

\setlength\LTleft{-2.85cm}
\begin{longtable}{|c|c|c|c|c|c|c|c|}
\caption{Searched SUGRA models} \label{tab:SUGRAscan}\\
\hline
\cellcolor[gray]{0.9} Model & \cellcolor[gray]{0.9} $K$ & \cellcolor[gray]{0.9} $W$ &\cellcolor[gray]{0.9} ICs & \cellcolor[gray]{0.9}$N_\textrm{end}$ &\cellcolor[gray]{0.9} $\omega_\textrm{end}^2$ & \cellcolor[gray]{0.9}Ref &  \cellcolor[gray]{0.9}Scanned range \\
\hline
\endfirsthead
\hline
\cellcolor[gray]{0.9} Model & \cellcolor[gray]{0.9} $K$ & \cellcolor[gray]{0.9} $W$ & \cellcolor[gray]{0.9}ICs & \cellcolor[gray]{0.9}$N_\textrm{end}$ & 
\cellcolor[gray]{0.9}$\omega_\textrm{end}^2$ & \cellcolor[gray]{0.9}Ref & \cellcolor[gray]{0.9}Scanned range \\
\hline
\endhead
\hline
\multicolumn{2}{r}{\footnotesize( To be continued)}
\endfoot
\bottomrule
\endlastfoot

EGNO\footnote{Note that these parameters are not limited to $p=0$, like the parameters presented in Section \ref{secEGNO}. They were found during a scan with \texttt{Inflation.jl} and \texttt{BlackBoxOptim.jl} as described above.} &
\eqref{Kegno}
& 
\eqref{Wegno} &
 $\begin{aligned}
\phi^1 &= 0.6496\\
\phi^2 &= 1.719\\
\pi^1 &= -0.2204\\
\pi^2 &= -0.6702\\
a &= 0.966\\
p &= 3.055\\
c &= 2258\\
\alpha &\equiv 1\\
S &\equiv 0
\end{aligned}$ &
 60.0 &
 10.58 & 
 \cite{EGNO} &
 $\begin{aligned}
\phi^1 &\in (0.0, 1.0)\\
\phi^2 &\in (0.0,5.0)\\
\pi^1 &\in (-\sqrt{2}, \sqrt{2})\\
\pi^2 &\in (-\sqrt{2}, \sqrt{2})\\
a &\in(0.0, 1.0)\\
p &\in (0,2\pi)\\
c &\in (0,3000) \end{aligned}$ \\ \hline
2 &
$\begin{aligned} &-3 \log[\Phi + \bar{\Phi}- S \bar{S}]\end{aligned}$
& 
$\begin{aligned} &\lambda - \mu \Phi \\&+ \beta S + \gamma \Phi S \end{aligned}$ &
 $\begin{aligned}
\phi^1 &= 14.59\\
\phi^2 &= 5.237\\
\phi^3 &= 19.75\\
\phi^4 &= 4.783\\
\pi^1 &= -0.8464\\
\pi^2 &= 0.8382\\
\pi^3 &= -0.5993\\
\pi^4 &= -0.4731\\
\lambda &= 0.0003386\\
\mu &= 0.00005\\
\beta &= 0.0003447\\
\gamma &= 0.0007987
\end{aligned}$ &
 60.0&
 $\sim10^{-9}$ & 
 \cite{Aldabergenov:2020atd} &
 $\begin{aligned}
\phi^i &\in (0.0, 20.0)\\
\lambda &\in (0.0,10^{-3})\\
\mu &\in (0.0,10^{-3})\\
\beta &\in (0.0,10^{-3})\\
\gamma &\in (0.0,10^{-3}) \end{aligned}$
 \\ \hline
 3 &
$\begin{aligned} \!-3 &\log[1 %\\
%&
+ \frac{1}{6}(\Phi - \bar{\Phi})^2\\
& \hskip1.4cm- \frac{1}{3} S \bar{S}]\end{aligned}$
& 
$\begin{aligned} &- S \left( \Phi^2 - \frac{v^2}{2} \right) \end{aligned}$ &
 $\begin{aligned}
\phi^1 &= 0.7162\\
\phi^2 &= 0.01539\\
\phi^3 &= 2.441\\
\phi^4 &= 0.005576\\
\pi^1 &= -1.026\\
\pi^2 &= -0.6852\\
\pi^3 &= 1.392\\
\pi^4 &= -0.9465\\
v &= 0.5917
\end{aligned}$ &
 60.0 &
 0.004 & 
 \cite{Kallosh:2010ug} &
 $\begin{aligned}
\phi^1 &\in (0.0, 5.0)\\
\phi^2 &\in (0.0, \sqrt{3})\\
\phi^3 &\in (0.0, \sqrt{6})\\
\phi^4 &\in (0.0, \sqrt{6})\\
\pi^i &\in (-\sqrt{2},\sqrt{2})\\
v &\in (0.0,20.0) \end{aligned}$
 \\ \hline
  4.1 &
$\begin{aligned} \frac{(\Phi+\bar{\Phi})^2}{2} + S \bar{S}\end{aligned}$
& 
$\begin{aligned}
S (1-e^{-\alpha \Phi})
 \end{aligned}$ &
 $\begin{aligned}
\phi^1 &= -0.04236\\
\phi^2 &= 0.4779\\
\phi^3 &= 9.098\\
\phi^4 &= 0.1057\\
\pi^1 &= -0.07004\\
\pi^2 &= -0.01171\\
\pi^3 &= 0.3973\\
\pi^4 &= 0.93\\
\alpha &= 0.4813
\end{aligned}$ &
 60.0 &
 0.00297 & 
 \cite{Kallosh:2014vja} &
 $\begin{aligned}
\phi^1 &\in (-2.0, 2.0)\\
\phi^2 &\in (0.0, 10.0)\\
\phi^3 &\in (0.0, 10.0)\\
\phi^4 &\in (0.0, 10.0)\\
\pi^i &\in (-\sqrt{2},\sqrt{2})\\
\alpha &\in (0.0,10.0) \end{aligned}$
 \\ \hline
   4.2 &
$\begin{aligned} -\frac{(\Phi+\bar{\Phi})^2}{2} + S \bar{S}
\end{aligned}$
& 
$\begin{aligned}
S \sin{\left(\frac{\alpha \Phi}{2}\right)}
 \end{aligned}$ &
 $\begin{aligned}
\phi^1 &= 1.921\\
\phi^2 &= 0.1958\\
\phi^3 &= 0.03938\\
\phi^4 &= 0.1432\\
\pi^1 &= 0.7423\\
\pi^2 &= 1.408\\
\pi^3 &= -0.04178\\
\pi^4 &= -0.6481\\
\alpha &= 2.107
\end{aligned}$ &
 13.27 &
 $\sim 10^{-8}$ & 
 \cite{Kallosh:2014vja} &
 $\begin{aligned}
\phi^1 &\in (-2.0, 2.0)\\
\phi^2 &\in (0.0, 10.0)\\
\phi^3 &\in (0.0, 10.0)\\
\phi^4 &\in (0.0, 10.0)\\
\pi^i &\in (-\sqrt{2},\sqrt{2})\\
\alpha &\in (0.0,10.0)
\end{aligned}$
 \\ \hline
    4.3 &
$\begin{aligned} -\frac{(\Phi+\bar{\Phi})^2}{2} + S \bar{S} \end{aligned}$
& 
$\begin{aligned}
S 
(&A \sin(\Phi \alpha/2) \\
+ &B \sin(\Phi \beta/2))
 \end{aligned}$ &
 $\begin{aligned}
\phi^1 &= 1.874\\
\phi^2 &= 0.7861\\
\phi^3 &= 0.09951\\
\phi^4 &= 0.8866\\
\pi^1 &= -0.5719\\
\pi^2 &= 1.136\\
\pi^3 &= 0.02645\\
\pi^4 &= 1.19\\
\alpha &= 1.325\\
\beta &= 4.175\\
A &= 3.718\\
B &= 1.402
\end{aligned}$ &
 60.0 &
 0.0025 & 
 \cite{Kallosh:2014vja} &
 $\begin{aligned}
\phi^1 &\in (-2.0, 2.0)\\
\phi^2 &\in (0.0, 10.0)\\
\phi^3 &\in (0.0, 10.0)\\
\phi^4 &\in (0.0, 10.0)\\
\pi^i &\in (-\sqrt{2},\sqrt{2})\\
\alpha &\in (0.0, 10.0)\\
\beta &\in (0.0, 10.0)\\
A &\in (0.0, 10.0)\\
B &\in (0.0, 10.0),
\end{aligned}$
 \\ \hline
  5 &
    $\begin{aligned}
-3 &\log{(\Phi+\bar{\Phi})} +\\
&- \log{(S+\bar{S})}
 \end{aligned}$ &
  $\begin{aligned}
-if &+ i h S \\
&+ i q \Phi
 \end{aligned}$ &
 $\begin{aligned}
\phi^1 &= 0.2574\\
\phi^2 &= 2.173\\
\phi^3 &= 1.203\\
\phi^4 &= 9.63\\
\pi^1 &= -0.2818\\
\pi^2 &= -0.01292\\
\pi^3 &= -1.37\\
\pi^4 &= -0.00025\\
h &= 1.7* 10^{-11}\\
q &= 3.625\\
f &= 2.599
\end{aligned}$ &
0.6489 &
 0.0014 & 
\cite{Blumenhagen:2015qda} &
 $\begin{aligned}
\phi^i &\in (0.0, 10.0)\\
\pi^i &\in (-\sqrt{2},\sqrt{2})\\
h &\in (0.0, 10.0)\\
q &\in (0.0, 10.0)\\
f &\in (0.0, 10.0)
\end{aligned}$
 \\ \hline
 6 &
   $\begin{aligned}
\Phi \bar{\Phi}
 \end{aligned}$ &
 $\begin{aligned}
\Phi - \frac{c+if}{\sqrt{2}}
 \end{aligned}$ &
 $\begin{aligned}
\phi^1 &= 0.9854\\
\phi^2 &= 0.09409\\
\pi^1 &= -0.2266\\
\pi^2 &= -0.6254\\
c &= 0.07026\\
f &= 0.1019
\end{aligned}$ &
 60.0 &
 $\sim10^{-5}$ & 
\cite{German:2019aoj} &
 $\begin{aligned}
\phi^1 &\in (0.0, 1.0)\\
\phi^2 &\in (0.0, 1.0)\\
\pi^i &\in (-\sqrt{2},\sqrt{2})\\
c &\in (0.0, 2.0)\\
f &\in (0.0, 2.0)
\end{aligned}$
 \\ \hline
   Poly$N$ &
    $\begin{aligned}
-3&\log(\Phi + \bar{\Phi}) \\
&+ S \bar{S}
 \end{aligned}$ &
  $\begin{aligned}
S\sum_{n=0}^N \frac{p_n\Phi^n}{n!}
 \end{aligned}$ &
 $\begin{aligned}
\phi^1 &= 1.777\\
\phi^2 &= 0.3429\\
\pi^1 &= -0.5227\\
\pi^2 &= -1.292\\
p_0 &= 93.3\\
p_1 &= 73.06\\
p_2 &= 31.9\\
p_3 &= 98.48\\
p_4 &= -3.389\\
p_5 &= 57.9\\
p_6 &= 69.24\\
p_7 &= -70.46\\
p_8 &= -85.69\\
p_9 &= -61.32\\
p_{10} &= -19.51\\
N &= 10\\
S &\equiv 0
\end{aligned}$ &
 60.0 &
 0.0033 & 
- &
 $\begin{aligned}
\phi^1 &\in (10^{-6},10^2)\\
\phi^2 &\in (-10^2,10^2)\\
\pi^i &\in (-\sqrt{2}, \sqrt{2})\\
p_i &\in (-10^2,10^2)\\
N &\in \{1,\ldots,10\}
\end{aligned}$
 \\ \hline
 QPoly$N$ &
    $\begin{aligned}
-3&\log(\Phi + \bar{\Phi}) \\
&+ S \bar{S}
 \end{aligned}$ &
  $\begin{aligned}
S\sum_{n=0}^N \frac{p_n\Phi^n}{\Phi^{N+1}n!}
 \end{aligned}$ &
 $\begin{aligned}
\phi^1 &= 1.992\\
\phi^2 &= -8.307\\
\pi^1 &= -1.233\\
\pi^2 &= 0.9123\\
p_0 &= -27.69\\
p_1 &= 33.45\\
p_2 &= 27.29\\
p_3 &= -93.18\\
p_4 &= -51.78\\
p_5 &= 23.04\\
p_6 &= 16.54\\
p_7 &= 67.45\\
p_8 &= 78.78\\
p_9 &= 38.83\\
p_{10} &= 58.3\\
N&=10\\
S&\equiv0
\end{aligned}$ &
60.0 &
 $\sim10^{-4}$ & 
- &
 $\begin{aligned}
\phi^1 &\in (10^{-6},10^2)\\
\phi^2 &\in (-10^2,10^2)\\
\pi^i &\in (-\sqrt{2}, \sqrt{2})\\
p_i &\in (-10^2,10^2)\\
N &\in \{1,\ldots,10\}
\end{aligned}$
 \\ \hline
 EPoly$N$ &
    $\begin{aligned}
-3&\log(\Phi + \bar{\Phi}) \\
&+ S \bar{S}
 \end{aligned}$ &
  $\begin{aligned}
S\sum_{n=0}^N \frac{p_n e^{n\Phi}}{n!}
 \end{aligned}$ &
 $\begin{aligned}
\phi^1 &= 0.5322\\
\phi^2 &= 15.81\\
\pi^1 &= -0.4712\\
\pi^2 &= 0.1758\\
p_0 &= -3.048\\
p_1 &= -41.88\\
p_2 &= -52.78\\
p_3 &= 97.38\\
p_4 &= 53.4\\
p_5 &= 75.42\\
p_6 &= 36.84\\
p_7 &= 73.26\\
p_8 &= 20.15\\
p_9 &= 32.92\\
p_{10} &= 36.91 \\
N&=10\\
S&\equiv0
\end{aligned}$ &
60.0 &
 0.54 & 
- &
 $\begin{aligned}
\phi^1 &\in (10^{-6},10^2)\\
\phi^2 &\in (-10^2,10^2)\\
\pi^i &\in (-\sqrt{2}, \sqrt{2})\\
p_i &\in (-10^2,10^2)\\
N &\in \{1,\ldots,10\}
\end{aligned}$
 \\ \hline
  Poly$N\alpha$\footnote{Although we present results for $N=1$, we searched up to $N=10$ without qualitatively different results. We also searched the similar models QPoly$N\alpha$ and EPoly$N\alpha$ (with this K\"ahler potential, but the same superpotentials as QPoly$N$ and EPoly$N$ respectively) and found a similarly large turning rate. As we understand from Section \ref{sec:noScaleModels}, the dynamics of this construction do not show a strong dependence on the structure of the superpotential in the rapid-turn regime.}
   &
    $\begin{aligned}
-3&\alpha\log(\Phi + \bar{\Phi}) \\
&+ S \bar{S}
 \end{aligned}$ &
  $\begin{aligned}
S\sum_{n=0}^N \frac{p_n\Phi^n}{n!}
 \end{aligned}$ &
 $\begin{aligned}
\phi^1 &= 1.735\\
\phi^2 &= -37.83\\
\pi^1 &= -0.4471\\
\pi^2 &= 0.2346\\
\alpha &= 0.001984\\
p_0 &= 69.23\\
p_1 &= -6.451\\
N&\equiv1\\
S&\equiv 0
\end{aligned}$ &
 60.0 &
 16.28 & 
- &
 $\begin{aligned}
\phi^1 &\in (10^{-6},10^2)\\
\phi^2 &\in (-10^2,10^2)\\
\pi^i &\in (-\sqrt{2}, \sqrt{2})\\
\alpha &\in (10^{-3},10.0)\\
p_i &\in (-10^2,10^2)
\end{aligned}$
 \\ \hline
Mono$N\alpha$\footnote{Note that for this example, the quoted values are barely inflationary, with an $\epsilon > 0.95$.} &
    $\begin{aligned}
-3\alpha\log(\Phi + \bar{\Phi})
 \end{aligned}$ &
  $\begin{aligned}
\Phi^N
 \end{aligned}$ &
 $\begin{aligned}
\phi^1 &= 0.9469\\
\phi^2 &= 10.09\\
\pi^1 &= -0.06989\\
\pi^2 &= -1.065\\
\alpha &= 0.9966\\
N &= 9.976
\end{aligned}$ &
 60.0 &
 1.294 & 
- &
 $\begin{aligned}
\phi^1 &\in (10^{-3},10^2)\\
\phi^2 &\in (0.0,10^2)\\
\pi^i &\in (-\sqrt{2}, \sqrt{2})\\
\alpha &\in (10^{-3},1.0)\\
N &\in (2.0,10.0)
\end{aligned}$
 \\ \hline
 Racetrack\footnote{Note that this is the original racetrack model without an uplifting term, which was included in \cite{Blanco-Pillado:2004aap}.} &
    $\begin{aligned}
    -3\alpha\log{(\Phi+\bar{\Phi})}
 \end{aligned}$ &
  $\begin{aligned}
  A e^{-a\Phi} + e^{-b\Phi}
 \end{aligned}$ &
 $\begin{aligned}
\phi^1 &= 0.8386\\
\phi^2 &= 64.83\\
\pi^1 &= 0.1642\\
\pi^2 &= 0.8657\\
\alpha &= 0.9097\\
a &= 0.1189\\
b &= 9.857\\
A &= 7.964
\end{aligned}$ &
 17.74 &
 $\sim10^{-8}$ & 
\cite{Blanco-Pillado:2004aap} &
 $\begin{aligned}
\phi^1 &\in (0.01,1.0) \\
\phi^2 &\in (0.0,100.0)\\
\pi^i &\in (\sqrt{2},\sqrt{2})\\
\alpha &\in (10^{-3},1.0)\\
a &\in (0.0,10.0)\\
b &\in (0.0,10.0)\\
A &\in (0.0,10.0)
\end{aligned}$
 \\ \hline
\end{longtable}

\end{appendix}

\bibliographystyle{JHEP}
\bibliography{refsACPRZ}

\providecommand{\href}[2]{#2}\begingroup\raggedright\begin{thebibliography}{10}

\bibitem{2020}
{\scshape Planck} collaboration, \emph{{Planck 2018 results. VI. Cosmological
  parameters}},
  \href{https://doi.org/10.1051/0004-6361/201833910}{\emph{Astron. Astrophys.}
  {\bfseries 641} (2020) A6}
  [\href{https://arxiv.org/abs/1807.06209}{{\ttfamily 1807.06209}}].

\bibitem{Hetz:2016ics}
A.~Hetz and G.A.~Palma, \emph{{Sound Speed of Primordial Fluctuations in
  Supergravity Inflation}},
  \href{https://doi.org/10.1103/PhysRevLett.117.101301}{\emph{Phys. Rev. Lett.}
  {\bfseries 117} (2016) 101301}
  [\href{https://arxiv.org/abs/1601.05457}{{\ttfamily 1601.05457}}].

\bibitem{Achucarro:2018vey}
A.~Ach\'ucarro and G.A.~Palma, \emph{{The string swampland constraints require
  multi-field inflation}},
  \href{https://doi.org/10.1088/1475-7516/2019/02/041}{\emph{JCAP} {\bfseries
  02} (2019) 041} [\href{https://arxiv.org/abs/1807.04390}{{\ttfamily
  1807.04390}}].

\bibitem{Obied:2018sgi}
G.~Obied, H.~Ooguri, L.~Spodyneiko and C.~Vafa, \emph{{De Sitter Space and the
  Swampland}},  \href{https://arxiv.org/abs/1806.08362}{{\ttfamily
  1806.08362}}.

\bibitem{Garg:2018reu}
S.K.~Garg and C.~Krishnan, \emph{{Bounds on Slow Roll and the de Sitter
  Swampland}}, \href{https://doi.org/10.1007/JHEP11(2019)075}{\emph{JHEP}
  {\bfseries 11} (2019) 075}
  [\href{https://arxiv.org/abs/1807.05193}{{\ttfamily 1807.05193}}].

\bibitem{Ooguri:2018wrx}
H.~Ooguri, E.~Palti, G.~Shiu and C.~Vafa, \emph{{Distance and de Sitter
  Conjectures on the Swampland}},
  \href{https://doi.org/10.1016/j.physletb.2018.11.018}{\emph{Phys. Lett. B}
  {\bfseries 788} (2019) 180}
  [\href{https://arxiv.org/abs/1810.05506}{{\ttfamily 1810.05506}}].

\bibitem{Chakraborty:2019dfh}
D.~Chakraborty, R.~Chiovoloni, O.~Loaiza-Brito, G.~Niz and I.~Zavala,
  \emph{{Fat inflatons, large turns and the $\eta$-problem}},
  \href{https://doi.org/10.1088/1475-7516/2020/01/020}{\emph{JCAP} {\bfseries
  01} (2020) 020} [\href{https://arxiv.org/abs/1908.09797}{{\ttfamily
  1908.09797}}].

\bibitem{Kaiser:2013sna}
D.I.~Kaiser and E.I.~Sfakianakis, \emph{{Multifield Inflation after Planck: The
  Case for Nonminimal Couplings}},
  \href{https://doi.org/10.1103/PhysRevLett.112.011302}{\emph{Phys. Rev. Lett.}
  {\bfseries 112} (2014) 011302}
  [\href{https://arxiv.org/abs/1304.0363}{{\ttfamily 1304.0363}}].

\bibitem{Christodoulidis:2018qdw}
P.~Christodoulidis, D.~Roest and E.I.~Sfakianakis, \emph{{Angular inflation in
  multi-field $\alpha$-attractors}},
  \href{https://doi.org/10.1088/1475-7516/2019/11/002}{\emph{JCAP} {\bfseries
  11} (2019) 002} [\href{https://arxiv.org/abs/1803.09841}{{\ttfamily
  1803.09841}}].

\bibitem{Christodoulidis:2019hhq}
P.~Christodoulidis, D.~Roest and R.~Rosati, \emph{{Many-field Inflation:
  Universality or Prior Dependence?}},
  \href{https://doi.org/10.1088/1475-7516/2020/04/021}{\emph{JCAP} {\bfseries
  04} (2020) 021} [\href{https://arxiv.org/abs/1907.08095}{{\ttfamily
  1907.08095}}].

\bibitem{Christodoulidis:2019jsx}
P.~Christodoulidis, D.~Roest and E.I.~Sfakianakis, \emph{{Scaling attractors in
  multi-field inflation}},
  \href{https://doi.org/10.1088/1475-7516/2019/12/059}{\emph{JCAP} {\bfseries
  12} (2019) 059} [\href{https://arxiv.org/abs/1903.06116}{{\ttfamily
  1903.06116}}].

\bibitem{Chen:2018brw}
X.~Chen, G.A.~Palma, B.~Scheihing~Hitschfeld and S.~Sypsas,
  \emph{{Reconstructing the Inflationary Landscape with Cosmological Data}},
  \href{https://doi.org/10.1103/PhysRevLett.121.161302}{\emph{Phys. Rev. Lett.}
  {\bfseries 121} (2018) 161302}
  [\href{https://arxiv.org/abs/1806.05202}{{\ttfamily 1806.05202}}].

\bibitem{Chen:2018uul}
X.~Chen, G.A.~Palma, W.~Riquelme, B.~Scheihing~Hitschfeld and S.~Sypsas,
  \emph{{Landscape tomography through primordial non-Gaussianity}},
  \href{https://doi.org/10.1103/PhysRevD.98.083528}{\emph{Phys. Rev. D}
  {\bfseries 98} (2018) 083528}
  [\href{https://arxiv.org/abs/1804.07315}{{\ttfamily 1804.07315}}].

\bibitem{Slosar:2019gvt}
A.~Slosar et~al., \emph{{Scratches from the Past: Inflationary Archaeology
  through Features in the Power Spectrum of Primordial Fluctuations}},
  \href{https://arxiv.org/abs/1903.09883}{{\ttfamily 1903.09883}}.

\bibitem{Braglia:2021ckn}
M.~Braglia, X.~Chen and D.K.~Hazra, \emph{{Comparing multi-field primordial
  feature models with the Planck data}},
  \href{https://doi.org/10.1088/1475-7516/2021/06/005}{\emph{JCAP} {\bfseries
  06} (2021) 005} [\href{https://arxiv.org/abs/2103.03025}{{\ttfamily
  2103.03025}}].

\bibitem{Braglia:2021rej}
M.~Braglia, X.~Chen and D.K.~Hazra, \emph{{Primordial Standard Clock Models and
  CMB Residual Anomalies}},  \href{https://arxiv.org/abs/2108.10110}{{\ttfamily
  2108.10110}}.

\bibitem{Fumagalli:2020adf}
J.~Fumagalli, S.~Renaux-Petel, J.W.~Ronayne and L.T.~Witkowski, \emph{{Turning
  in the landscape: a new mechanism for generating Primordial Black Holes}},
  \href{https://arxiv.org/abs/2004.08369}{{\ttfamily 2004.08369}}.

\bibitem{Braglia:2020eai}
M.~Braglia, D.K.~Hazra, F.~Finelli, G.F.~Smoot, L.~Sriramkumar and
  A.A.~Starobinsky, \emph{{Generating PBHs and small-scale GWs in two-field
  models of inflation}},
  \href{https://doi.org/10.1088/1475-7516/2020/08/001}{\emph{JCAP} {\bfseries
  08} (2020) 001} [\href{https://arxiv.org/abs/2005.02895}{{\ttfamily
  2005.02895}}].

\bibitem{Palma:2020ejf}
G.A.~Palma, S.~Sypsas and C.~Zenteno, \emph{{Seeding primordial black holes in
  multifield inflation}},
  \href{https://doi.org/10.1103/PhysRevLett.125.121301}{\emph{Phys. Rev. Lett.}
  {\bfseries 125} (2020) 121301}
  [\href{https://arxiv.org/abs/2004.06106}{{\ttfamily 2004.06106}}].

\bibitem{Anguelova:2020nzl}
L.~Anguelova, \emph{{On Primordial Black Holes from Rapid Turns in Two-field
  Models}}, \href{https://doi.org/10.1088/1475-7516/2021/06/004}{\emph{JCAP}
  {\bfseries 06} (2021) 004}
  [\href{https://arxiv.org/abs/2012.03705}{{\ttfamily 2012.03705}}].

\bibitem{Barausse:2020rsu}
E.~Barausse et~al., \emph{{Prospects for Fundamental Physics with LISA}},
  \href{https://doi.org/10.1007/s10714-020-02691-1}{\emph{Gen. Rel. Grav.}
  {\bfseries 52} (2020) 81} [\href{https://arxiv.org/abs/2001.09793}{{\ttfamily
  2001.09793}}].

\bibitem{Fumagalli:2020nvq}
J.~Fumagalli, S.~Renaux-Petel and L.T.~Witkowski, \emph{{Oscillations in the
  stochastic gravitational wave background from sharp features and particle
  production during inflation}},
  \href{https://doi.org/10.1088/1475-7516/2021/08/030}{\emph{JCAP} {\bfseries
  08} (2021) 030} [\href{https://arxiv.org/abs/2012.02761}{{\ttfamily
  2012.02761}}].

\bibitem{Fumagalli:2021cel}
J.~Fumagalli, S.e.~Renaux-Petel and L.T.~Witkowski, \emph{{Resonant features in
  the stochastic gravitational wave background}},
  \href{https://doi.org/10.1088/1475-7516/2021/08/059}{\emph{JCAP} {\bfseries
  08} (2021) 059} [\href{https://arxiv.org/abs/2105.06481}{{\ttfamily
  2105.06481}}].

\bibitem{Domenech:2021ztg}
G.~Dom\`enech, \emph{{Scalar induced gravitational waves review}},
  \href{https://arxiv.org/abs/2109.01398}{{\ttfamily 2109.01398}}.

\bibitem{Bjorkmo:2019fls}
T.~Bjorkmo, \emph{{Rapid-Turn Inflationary Attractors}},
  \href{https://doi.org/10.1103/PhysRevLett.122.251301}{\emph{Phys. Rev. Lett.}
  {\bfseries 122} (2019) 251301}
  [\href{https://arxiv.org/abs/1902.10529}{{\ttfamily 1902.10529}}].

\bibitem{Bjorkmo:2019aev}
T.~Bjorkmo and M.D.~Marsh, \emph{{Hyperinflation generalised: from its
  attractor mechanism to its tension with the \textquoteleft{}swampland
  conditions\textquoteright{}}},
  \href{https://doi.org/10.1007/JHEP04(2019)172}{\emph{JHEP} {\bfseries 04}
  (2019) 172} [\href{https://arxiv.org/abs/1901.08603}{{\ttfamily
  1901.08603}}].

\bibitem{Pinol:2020kvw}
L.~Pinol, \emph{{Multifield inflation beyond $N_\mathrm{field}=2$:
  non-Gaussianities and single-field effective theory}},
  \href{https://doi.org/10.1088/1475-7516/2021/04/002}{\emph{JCAP} {\bfseries
  04} (2021) 002} [\href{https://arxiv.org/abs/2011.05930}{{\ttfamily
  2011.05930}}].

\bibitem{Aragam:2020uqi}
V.~Aragam, S.~Paban and R.~Rosati, \emph{{The Multi-Field, Rapid-Turn
  Inflationary Solution}},
  \href{https://doi.org/10.1007/JHEP03(2021)009}{\emph{JHEP} {\bfseries 03}
  (2021) 009} [\href{https://arxiv.org/abs/2010.15933}{{\ttfamily
  2010.15933}}].

\bibitem{Achucarro:2019mea}
A.~Ach\'ucarro and Y.~Welling, \emph{{Orbital Inflation: inflating along an
  angular isometry of field space}},
  \href{https://arxiv.org/abs/1907.02020}{{\ttfamily 1907.02020}}.

\bibitem{Aragam:2019omo}
V.~Aragam, S.~Paban and R.~Rosati, \emph{{Multi-field Inflation in High-Slope
  Potentials}},
  \href{https://doi.org/10.1088/1475-7516/2020/04/022}{\emph{JCAP} {\bfseries
  04} (2020) 022} [\href{https://arxiv.org/abs/1905.07495}{{\ttfamily
  1905.07495}}].

\bibitem{Achucarro:2019pux}
A.~Ach\'ucarro, E.J.~Copeland, O.~Iarygina, G.A.~Palma, D.-G.~Wang and
  Y.~Welling, \emph{{Shift-symmetric orbital inflation: Single field or
  multifield?}}, \href{https://doi.org/10.1103/PhysRevD.102.021302}{\emph{Phys.
  Rev. D} {\bfseries 102} (2020) 021302}
  [\href{https://arxiv.org/abs/1901.03657}{{\ttfamily 1901.03657}}].

\bibitem{Starobinsky:1980te}
A.A.~Starobinsky, \emph{{A New Type of Isotropic Cosmological Models Without
  Singularity}},
  \href{https://doi.org/10.1016/0370-2693(80)90670-X}{\emph{Phys. Lett. B}
  {\bfseries 91} (1980) 99}.

\bibitem{Christodoulidis:2019mkj}
P.~Christodoulidis, D.~Roest and E.I.~Sfakianakis, \emph{{Attractors,
  Bifurcations and Curvature in Multi-field Inflation}},
  \href{https://doi.org/10.1088/1475-7516/2020/08/006}{\emph{JCAP} {\bfseries
  08} (2020) 006} [\href{https://arxiv.org/abs/1903.03513}{{\ttfamily
  1903.03513}}].

\bibitem{Achucarro:2010da}
A.~Achucarro, J.-O.~Gong, S.~Hardeman, G.A.~Palma and S.P.~Patil,
  \emph{{Features of heavy physics in the CMB power spectrum}},
  \href{https://doi.org/10.1088/1475-7516/2011/01/030}{\emph{JCAP} {\bfseries
  01} (2011) 030} [\href{https://arxiv.org/abs/1010.3693}{{\ttfamily
  1010.3693}}].

\bibitem{KLR}
R.~Kallosh, A.~Linde and T.~Rube, \emph{{General inflaton potentials in
  supergravity}}, \href{https://doi.org/10.1103/PhysRevD.83.043507}{\emph{Phys.
  Rev. D} {\bfseries 83} (2011) 043507}
  [\href{https://arxiv.org/abs/1011.5945}{{\ttfamily 1011.5945}}].

\bibitem{RSZ}
D.~Roest, M.~Scalisi and I.~Zavala, \emph{{K\"ahler potentials for Planck
  inflation}}, \href{https://doi.org/10.1088/1475-7516/2013/11/007}{\emph{JCAP}
  {\bfseries 11} (2013) 007} [\href{https://arxiv.org/abs/1307.4343}{{\ttfamily
  1307.4343}}].

\bibitem{Garcia-Saenz:2018ifx}
S.~Garcia-Saenz, S.~Renaux-Petel and J.~Ronayne, \emph{{Primordial fluctuations
  and non-Gaussianities in sidetracked inflation}},
  \href{https://doi.org/10.1088/1475-7516/2018/07/057}{\emph{JCAP} {\bfseries
  07} (2018) 057} [\href{https://arxiv.org/abs/1804.11279}{{\ttfamily
  1804.11279}}].

\bibitem{Ketov:2021fww}
S.V.~Ketov, \emph{{Multi-Field versus Single-Field in the Supergravity Models
  of Inflation and Primordial Black Holes}},
  \href{https://doi.org/10.3390/universe7050115}{\emph{Universe} {\bfseries 7}
  (2021) 115}.

\bibitem{Brown:2017osf}
A.R.~Brown, \emph{{Hyperbolic Inflation}},
  \href{https://doi.org/10.1103/PhysRevLett.121.251601}{\emph{Phys. Rev. Lett.}
  {\bfseries 121} (2018) 251601}
  [\href{https://arxiv.org/abs/1705.03023}{{\ttfamily 1705.03023}}].

\bibitem{AlvarezGaume:2011xv}
L.~Alvarez-Gaume, C.~Gomez and R.~Jimenez, \emph{{A Minimal Inflation
  Scenario}}, \href{https://doi.org/10.1088/1475-7516/2011/03/027}{\emph{JCAP}
  {\bfseries 03} (2011) 027} [\href{https://arxiv.org/abs/1101.4948}{{\ttfamily
  1101.4948}}].

\bibitem{Achucarro:2012hg}
A.~Achucarro, S.~Mooij, P.~Ortiz and M.~Postma, \emph{{Sgoldstino inflation}},
  \href{https://doi.org/10.1088/1475-7516/2012/08/013}{\emph{JCAP} {\bfseries
  08} (2012) 013} [\href{https://arxiv.org/abs/1203.1907}{{\ttfamily
  1203.1907}}].

\bibitem{Ferrara:2014kva}
S.~Ferrara, R.~Kallosh and A.~Linde, \emph{{Cosmology with Nilpotent
  Superfields}}, \href{https://doi.org/10.1007/JHEP10(2014)143}{\emph{JHEP}
  {\bfseries 10} (2014) 143} [\href{https://arxiv.org/abs/1408.4096}{{\ttfamily
  1408.4096}}].

\bibitem{Cremmer:1983bf}
E.~Cremmer, S.~Ferrara, C.~Kounnas and D.V.~Nanopoulos, \emph{{Naturally
  Vanishing Cosmological Constant in N=1 Supergravity}},
  \href{https://doi.org/10.1016/0370-2693(83)90106-5}{\emph{Phys. Lett. B}
  {\bfseries 133} (1983) 61}.

\bibitem{EGNO}
J.~Ellis, M.A.G.~Garcia, D.V.~Nanopoulos and K.A.~Olive, \emph{{A No-Scale
  Inflationary Model to Fit Them All}},
  \href{https://doi.org/10.1088/1475-7516/2014/08/044}{\emph{JCAP} {\bfseries
  08} (2014) 044} [\href{https://arxiv.org/abs/1405.0271}{{\ttfamily
  1405.0271}}].

\bibitem{Calderon-Infante:2020dhm}
J.~Calder\'on-Infante, A.M.~Uranga and I.~Valenzuela, \emph{{The Convex Hull
  Swampland Distance Conjecture and Bounds on Non-geodesics}},
  \href{https://doi.org/10.1007/JHEP03(2021)299}{\emph{JHEP} {\bfseries 03}
  (2021) 299} [\href{https://arxiv.org/abs/2012.00034}{{\ttfamily
  2012.00034}}].

\bibitem{Conlon:2005jm}
J.P.~Conlon and F.~Quevedo, \emph{{Kahler moduli inflation}},
  \href{https://doi.org/10.1088/1126-6708/2006/01/146}{\emph{JHEP} {\bfseries
  01} (2006) 146} [\href{https://arxiv.org/abs/hep-th/0509012}{{\ttfamily
  hep-th/0509012}}].

\bibitem{Bond:2006nc}
J.R.~Bond, L.~Kofman, S.~Prokushkin and P.M.~Vaudrevange, \emph{{Roulette
  inflation with Kahler moduli and their axions}},
  \href{https://doi.org/10.1103/PhysRevD.75.123511}{\emph{Phys. Rev. D}
  {\bfseries 75} (2007) 123511}
  [\href{https://arxiv.org/abs/hep-th/0612197}{{\ttfamily hep-th/0612197}}].

\bibitem{Blanco-Pillado:2009dmu}
J.J.~Blanco-Pillado, D.~Buck, E.J.~Copeland, M.~Gomez-Reino and N.J.~Nunes,
  \emph{{Kahler Moduli Inflation Revisited}},
  \href{https://doi.org/10.1007/JHEP01(2010)081}{\emph{JHEP} {\bfseries 01}
  (2010) 081} [\href{https://arxiv.org/abs/0906.3711}{{\ttfamily 0906.3711}}].

\bibitem{Andriot:2021rdy}
D.~Andriot, \emph{{Tachyonic de Sitter solutions of 10d type II
  supergravities}},  \href{https://arxiv.org/abs/2101.06251}{{\ttfamily
  2101.06251}}.

\bibitem{Kallosh:2010ug}
R.~Kallosh and A.~Linde, \emph{{New models of chaotic inflation in
  supergravity}},
  \href{https://doi.org/10.1088/1475-7516/2010/11/011}{\emph{JCAP} {\bfseries
  11} (2010) 011} [\href{https://arxiv.org/abs/1008.3375}{{\ttfamily
  1008.3375}}].

\bibitem{Kallosh:2014vja}
R.~Kallosh, A.~Linde and B.~Vercnocke, \emph{{Natural Inflation in Supergravity
  and Beyond}}, \href{https://doi.org/10.1103/PhysRevD.90.041303}{\emph{Phys.
  Rev. D} {\bfseries 90} (2014) 041303}
  [\href{https://arxiv.org/abs/1404.6244}{{\ttfamily 1404.6244}}].

\bibitem{German:2019aoj}
G.~Germ\'an, J.C.~Hidalgo, F.X.~Linares Cede\~no, A.~Montiel and
  J.A.~V\'azquez, \emph{{Simple supergravity model of inflation constrained
  with Planck 2018 data}},
  \href{https://doi.org/10.1103/PhysRevD.101.023507}{\emph{Phys. Rev. D}
  {\bfseries 101} (2020) 023507}
  [\href{https://arxiv.org/abs/1909.02019}{{\ttfamily 1909.02019}}].

\bibitem{Blumenhagen:2015qda}
R.~Blumenhagen, A.~Font, M.~Fuchs, D.~Herschmann and E.~Plauschinn,
  \emph{{Towards Axionic Starobinsky-like Inflation in String Theory}},
  \href{https://doi.org/10.1016/j.physletb.2015.05.001}{\emph{Phys. Lett. B}
  {\bfseries 746} (2015) 217}
  [\href{https://arxiv.org/abs/1503.01607}{{\ttfamily 1503.01607}}].

\bibitem{Aldabergenov:2020atd}
Y.~Aldabergenov, \emph{{Volkov\textendash{}Akulov\textendash{}Starobinsky
  supergravity revisited}},
  \href{https://doi.org/10.1140/epjc/s10052-020-7888-8}{\emph{Eur. Phys. J. C}
  {\bfseries 80} (2020) 329}
  [\href{https://arxiv.org/abs/2001.06617}{{\ttfamily 2001.06617}}].

\bibitem{Blanco-Pillado:2004aap}
J.J.~Blanco-Pillado, C.P.~Burgess, J.M.~Cline, C.~Escoda, M.~Gomez-Reino,
  R.~Kallosh et~al., \emph{{Racetrack inflation}},
  \href{https://doi.org/10.1088/1126-6708/2004/11/063}{\emph{JHEP} {\bfseries
  11} (2004) 063} [\href{https://arxiv.org/abs/hep-th/0406230}{{\ttfamily
  hep-th/0406230}}].

\bibitem{Inflationjl}
R.~Rosati, \emph{Inflation.jl -- a julia package for numerical evaluation of
  cosmic inflation models using the transport method},  July, 2020.
\newblock 10.5281/zenodo.4708348.

\end{thebibliography}\endgroup

\end{document}